\documentclass[journal,transmag]{IEEEtran}
\usepackage{cite}
\usepackage{amsmath,amssymb,amsfonts}
\usepackage{algorithmic}
\usepackage{graphicx}
\usepackage{textcomp}
\usepackage{multirow}
\usepackage{bm}
\usepackage{hyperref}
\usepackage{color}
\usepackage{rotating}

\begin{document}
\title{{\color{black}Deep-learning-based Optimization of the \\ Under-sampling Pattern in MRI}}

\author{Cagla D. Bahadir*, Alan Q. Wang*, Adrian V. Dalca, and Mert R. Sabuncu \thanks{Cagla D. Bahadir and Mert R. Sabuncu are with the Meinig School of Biomedical Engineering, Cornell University. Alan Q. Wang and Mert R. Sabuncu are with the School of Electrical and Computer Engineering, Cornell University. Adrian V. Dalca is with the Computer Science and Artificial Intelligence Lab at the Massachusetts Institute of Technology and the A.A. Martinos Center for Biomedical Imaging at the Massachusetts General Hospital.
{\color{black}* indicates equal contribution. Cagla D. Bahadir and Alan Q. Wang are co-first authors.}
}}

\maketitle

\begin{abstract}
In compressed sensing MRI (CS-MRI), k-space measurements are under-sampled to achieve accelerated scan times. 
CS-MRI presents two fundamental problems: (1) where to sample and (2) how to reconstruct an under-sampled scan.
In this paper, we tackle both problems simultaneously for the specific case of 2D Cartesian sampling, using a novel end-to-end learning framework that we call LOUPE (Learning-based Optimization of the Under-sampling PattErn).
Our method trains a neural network model on a set of full-resolution MRI scans, which are retrospectively under-sampled {\color{black}on a 2D Cartesian grid} and forwarded to an anti-aliasing {\color{black}(a.k.a. reconstruction)} model that computes a reconstruction, which is in turn compared with the input.
This formulation enables a data-driven optimized under-sampling pattern at a given sparsity level.
In our experiments, we demonstrate that LOUPE-optimized under-sampling masks are data-dependent, varying significantly with the imaged anatomy, and perform well with different reconstruction methods.
We present empirical results obtained with a large-scale, publicly available knee MRI dataset, where LOUPE offered superior reconstruction quality across different conditions.
Even with an aggressive 8-fold acceleration rate, LOUPE's reconstructions contained much of the anatomical detail that was missed by alternative masks and reconstruction methods.
Our experiments also show how LOUPE yielded optimal under-sampling patterns that were significantly different for brain vs knee MRI scans.
Our code is made freely available at \url{https://github.com/cagladbahadir/LOUPE/}.
\end{abstract}


\section{Introduction}
\label{sec:introduction}
\IEEEPARstart{M}{agnetic} Resonance Imaging (MRI) is a ubiquitous, non-invasive, and versatile biomedical imaging technology. 
A central challenge in MRI is long scan times, which constrains accessibility and leads to high costs. 
One remedy is to accelerate MRI via compressed sensing~\cite{lustig2008compressed,gamper2008compressed}.
In compressed sensing MRI, k-space data (i.e., the Fourier transform of the image) is sampled below the Nyquist-Shannon rate~\cite{lustig2008compressed}, which is often referred to as ``under-sampling.''
Given an under-sampled set of measurements, the objective is to ``reconstruct'' the full-resolution MRI.
This is the main problem that most of the compressed sensing literature is focused on and is conventionally formulated as an optimization problem that trades off two objectives: one that quantifies the fit between the measurements and the reconstruction (sometimes referred to as data consistency), and another that captures prior knowledge on the distribution of MRI data. 
This latter objective is often achieved via the incorporation of regularization terms, such as the total variation penalty and/or a sparsity-inducing norm on transformation coefficients, like wavelets or a dictionary decomposition~\cite{ma2008efficient}. 
Such regularization functions aim to capture different properties of real-world MR images, which in turn help the ill-posed  reconstruction problem by guiding to more realistic solutions. 
One approach to develop a data-driven regularization function is to construct a sparsifying dictionary, for example, based on image patches~\cite{qu2014magnetic, ravishankar2011mr, zhan2016fast}.

Once the optimization problem is set up, the reconstruction algorithm often iteratively minimizes the regularization term(s) and enforces data consistency in k-space. This is classically solved for each acquired dataset, independently, and from scratch - a process that can be computationally demanding. 
As we describe in the following section, there has been a recent surge in machine learning based methods that take a different, and often computationally more efficient approach to solving the reconstruction problem.
These techniques are gradually becoming more widespread and expected to complement existing regularized optimization based approaches.

Another critical component of compressed sensing MRI is the under-sampling pattern. 
For a given acceleration rate, there is an exponentially large number of possible patterns one can implement for under-sampling. 
Each of these under-sampling patterns will in general lead to different reconstruction performance that will depend on the statistics of the data and the utilized reconstruction method. 
One way to view this is to regard the reconstruction model as imputing the parts of the Fourier spectrum that was not sampled.
For example, a frequency component that is constant for all possible datasets we might observe, does not need to be sampled when we use a reconstruction model that can leverage this information.
On the other hand, highly variable parts of the spectrum will likely need to be measured to achieve accurate reconstructions.
This simple viewpoint ignores the potentially multi-variate nature of the distribution of k-space measurements.
For instance, missing parts of the Fourier spectrum might be reliably imputed from other acquired parts, due to strong statistical dependency.
Widely used under-sampling strategies in compressed sensing MRI include Random Uniform~\cite{gamper2008compressed}, Variable Density~\cite{wang2010variable} and equi-spaced Cartesian~\cite{haldar2011compressed} with skipped lines. 
These techniques, however, are often implemented heuristically, not in a data-driven adaptive fashion, and their popularity is largely due to their ease of implementation and good performance when coupled with popular reconstruction methods.
As we discuss below, some recent efforts compute optimized under-sampling patterns - a literature that is closely related to our primary objective.

In this paper, we propose to merge the two core problems of compressed sensing MRI: (1) optimizing the under-sampling pattern and (2) reconstruction.
We consider these two problems simultaneously because they are closely inter-related.
Specifically, the optimal under-sampling pattern should, in general, depend on the reconstruction method and vice versa.
Inspired by recent developments in machine learning, we propose a novel end-to-end deep learning strategy to solve the combined problem.
We call our method LOUPE, which stands for Learning-based Optimization of the Under-sampling PattErn.
A preliminary version of LOUPE was published as a conference paper~\cite{bahadir2019learning}. 
In this journal paper, we present an extended treatment of the literature, an important modification to the LOUPE objective that enables us to directly set the desired sparsity level while obviating the need to identify the value of an extra hyper-parameter, more details on the methods, and new experimental results.

The rest of the paper is organized as follows.
In the following two sections, we review two closely related bodies of work: a rapidly growing list of recent papers that use machine learning for efficient and accurate reconstruction; and several proposed approaches to optimize the under-sampling pattern.
Section~\ref{sec:method} then presents the mathematical and implementation details of the proposed method, LOUPE. 
Section~\ref{sec:experiments} presents our experiments, where we compare the performance obtained with several benchmark reconstruction methods and under-sampling patterns. 
Section~\ref{sec:conc} concludes with a discussion.

\section{Machine Learning for Under-sampled Image Reconstruction}
\label{sec:ML-Recon}

Over the last few years, machine learning methods have been increasingly used for medical image reconstruction~\cite{sun2016deep,tezcan2018mr,wang2018image,zhu2018image,aggarwal2018modl,hammernik2018learning}, including compressed sensing MRI. 
An earlier example is the Bayesian non-parametric dictionary learning approach~\cite{huang2014bayesian}. 
In this framework, dictionary learning is used in the reconstruction problem to obtain a regularization objective that is customized to the image at hand. The reconstruction problem is solved via a traditional optimization approach.

More recently, fueled by the success of deep learning, several machine learning techniques have been proposed to implement efficient and accurate reconstruction models.
For instance, in a high profile paper, Zhu et al. demonstrated a versatile supervised learning approach, called AUTOMAP~\cite{zhu2018image}. This method uses a neural network model to map sensor measurements directly to reconstruction images, via a manifold representation. 
The experimental results show that AUTOMAP can produce high quality reconstructions that are robust to noise and other artifacts.
Another machine learning based reconstruction approach involves using a neural network to efficiently execute the iterations of the Alternating Direction Method of Multipliers (ADMM) method that solves a conventional optimization problem~\cite{sun2016deep,yang2017admm}. This technique, called ADDM-Net, uses an ``unrolled'' neural network architecture to implement the iterative optimization procedure.
In a similar approach, an unrolled neural network is used to implement the iterations of the Landweber method~\cite{hammernik2018learning}.



Another class of methods rely on the U-Net architecture~\cite{ronneberger2015u} or its variants.
For example U-Net-like models have been trained in a supervised fashion to remove aliasing artifacts in the imaging domain~\cite{lee2017deep,hyun2018deep}.
The U-Net architecture has also been used by appending with a forward physics based model to learn phase masks for depth estimation in other computer vision applications ~\cite{wu2019phasecam3d}. 
In these methods, the neural network takes as input a poor quality reconstruction (e.g., obtained via a simple inverse Fourier transform applied to zero-filled k-space data) to compute a high quality output, where the objective is to minimize the loss function  (e.g., squared difference) between the output and the ground truth provided during training.
Inspired by the success of Generative Adversarial Networks (GANs)~\cite{goodfellow2014generative}, several groups have proposed the use of an adversarial loss, in addition to the more conventional loss functions, to obtain better quality reconstructions~\cite{yang2018dagan,quan2018compressed,mardani2017deep,lei2019wasserstein}. 


The method we propose in this paper builds on the neural net-based reconstruction framework, which offers a computationally efficient and differentiable anti-aliasing model.
We combine the anti-aliasing model with a retrospective under-sampling module, and learn both blocks simultaneously. 

\section{Data-driven Under-sampling in Compressed Sensing MRI}
\label{sec:Opt-Sub}

The under-sampling pattern in k-space is closely related to reconstruction performance. 
Several under-sampling patterns are widely used, including Random Uniform~\cite{gamper2008compressed}, Variable Density~\cite{lustig2007sparse,wang2010variable} and equi-spaced Cartesian~\cite{haldar2011compressed} with skipped lines. 
Since the early work by Lustig et al.~\cite{lustig2007sparse}, random or stochastic under-sampling strategies are commonly used, as they induce noise-like artifacts that are often easier to remove during reconstruction. 
However, as several papers have pointed out, the quality of reconstruction can also be improved by adapting the under-sampling pattern to the data and application.
We underscore that there is a related, but more general problem of optimizing projections in compressed sensing, which has received considerable attention~\cite{elad2007optimized,xu2010optimized,puy2011variable,li2013projection}. 
Below, we review adaptive under-sampling strategies in compressed sensing MRI, which is the focus of our paper. 

Knoll et al. proposed an under-sampling strategy that follows the power spectrum of a provided example (template) image~\cite{knoll2011adapted}. 
This method collects denser samples in parts of k-space that has most of the power concentrated, such as the low-frequency components of real-world MRIs. 
The authors argue that most of the anatomical variability is reflected in the phase and not magnitude of the Fourier spectrum, justifying the decision to rely on the magnitude to construct the under-sampling pattern.
In experiments, they apply their method to both knee and brain MRI scans, demonstrating better performance than parametric variable density under-sampling patterns~\cite{lustig2007sparse}.

Kumar Anand et al. presented a second-order cone optimization technique for finding the optimal under-sampling trajectory in volumetric MRI scans~\cite{kumar2008durga}. 
Their heuristic strategy accounts for coverage, hardware limitations and signal generation in a single convex optimization problem. 
Building on this approach, Curtis et al. employed a genetic algorithm to optimize sampling trajectories in k-space, while accounting for multi-coil configurations~\cite{curtis2008random}.

Seeger et al. employed a Bayesian inference approach to optimize Cartesian and spiral trajectories~\cite{seeger2010optimization}. 
Their iterative, greedy technique seeks the under-sampling pattern in k-space that minimizes the uncertainty in the computed reconstruction, conditioned on the acquired measurements. 
In a recent paper that explores a similar direction, Haldar et al. proposed a new framework called Oracle-based Experiment Design for Imaging Parsimoniously Under Sparsity constraints (OEDIPUS)~\cite{haldar2019oedipus}.
OEDIPUS solves an integer programming problem to minimize a lower (Cramer-Rao) bound on the variance of the reconstruction. 

Roman et al. offered a novel theoretical perspective on compressed sensing, which underscores the importance of adapting the under-sampling strategy to the structure in the data~\cite{roman2014asymptotic}.
The authors also presented an innovative multilevel sampling approach that depends on the resolution and the structure of the data that empirically outperforms competing under-sampling schemes. 


An alternative class of data-driven approaches aims to find the optimal under-sampling pattern that yields the best reconstruction or imputation quality, using some full-resolution training data that is retrospectively under-sampled~\cite{sherry2019learning}.
The central challenge in this framework is to efficiently solve two computationally challenging nested optimization problems.
The first, outer problem is to identify the optimal under-sampling mask or sampling trajectory that achieves best reconstruction quality, which is in turn the result of an inner optimization step.
Several authors have proposed to solve this nested pair of problems via heuristic, greedy algorithms~\cite{ravishankar2011adaptive,liu2012under,baldassarre2016learning, gozcu2018learning,sanchez2020scalable}.
In an empirical study, Zijlstra et al. demonstrated that a data-driven optimization approach~\cite{zijlstra2016evaluation,liu2012under} can yield better reconstructions than those obtained with more conventional methods~\cite{lustig2007sparse, knoll2011adapted}.
{\color{black} However, prior data-driven methods mostly lack the computational efficiency to handle large-scale full-resolution (training) data. 
Furthermore, they are often not flexible enough to deal with different types of under-sampling schemes.} 

In this paper, we present a novel, {\color{black} flexible, and computationally efficient} data-driven approach to optimize the under-sampling pattern.
Instead of formulating the problem as two nested optimization problems, we leverage modern machine learning techniques and take an end-to-end learning perspective.
Similar to recent deep learning based reconstruction techniques~\cite{lee2017deep,hyun2018deep}, our implementation employs the U-Net~\cite{ronneberger2015u} architecture, which we append with a probabilistic under-sampling step that is also learned. 
We assume we are provided with full-resolution MRI data, which we retrospectively under-sample. 
{\color{black}We note that there have been contemporaneous efforts~\cite{weiss2019learning,aggarwal2019joint,huijben2020learning} that build on our prior work~\cite{bahadir2019learning} to take a deep learning-based approach similar to ours for optimizing the under-sampling pattern in compressed sensing.}



\section{Method}
\label{sec:method}

\subsection{Learning-based Optimization of Under-sampling Pattern}

LOUPE solves the two problems of compressed sensing simultaneously: (1) identifying the optimal under-sampling pattern; and (2) reconstructing from the under-sampled measurements. 
Building on stochastic strategies of compressed sensing, we consider a probabilistic mask $\bm{P}$, which describes an independent Bernoulli random variable at each k-space point. 
The probabilistic mask
$\bm{P}$ is defined on the full-resolution k-space grid, and at each point takes on non-negative continuous probability values, i.e., {\color{black}$\bm{P} \in [0,1]^d$}, where $d$ is the total number of grid points.
For example, for a $100 \times 100$ image, $d = 10,000$.
{\color{black}We parameterize  $\bm{P}$ with an unconstrained image $\bm{O} \in \mathbb{R}^d$, such that $\bm{P} = \sigma_t(\bm{O})$.
Here, $\sigma_t$ denotes an element-wise sigmoid, with slope  $t$, which is treated as a hyper-parameter.
$\bm{P}_i = \frac{1}{1+e^{-t \bm{O}_i}}$, where the sub-script $i$ indexes the grid points.}
Binary realizations drawn from $\bm{P}$ represent an under-sampling mask $\bm{M} \in \{0,1\}^d$: i.e., $\bm{M} \sim \prod_{i=1}^d \mathcal{B}(\bm{P}_i)$, where $\mathcal{B}(p)$ denotes a Bernoulli random variable with parameter $p$.
The binary mask $\bm{M}$ has value 1 for grid points that are acquired and 0 for points that were not acquired. 

Suppose we are provided with a collection of complex-valued full-resolution images, denoted as $\{\bm{x}_j \in \mathbb{C}^d\}_{j=1}^{n}$,  where $j$ corresponds to the scan number and $n$ is the total number of scans.
In our following treatment, without loss of generality, we will assume the provided data are in image domain. 
This can be modified to accept raw k-space measurements, by simply removing a forward Fourier transform.
In the LOUPE framework, we solve the following problem:
%
{\color{black}
\begin{multline}
\min_{\bm{O}, \theta} \mathbb{E}_{\bm{M} \sim \prod_{i=1} \mathcal{B}(\sigma_t(\bm{O}_i))}
    \sum_{j=1}^n \| A_\theta (F^{H} \textrm{diag}(\bm{M}) F \bm{x}_j) - \bm{x}_j \|_2^2, \\
    {\textrm{such that } \frac{1}{d}\|\sigma_t(\bm{O})\|_1 = \alpha}
\label{eq:LOUPE1}
\end{multline}}
where $F \in \mathbb{C}^{d \times d}$ stands for the (forward) Fourier transform matrix, $F^{H}$ is the inverse Fourier transform matrix,  $A_\theta$ denotes an anti-aliasing reconstruction function parameterized with $\theta$, $\| \cdot \|_2$ denotes the L2 norm, $\| \cdot \|_1$ denotes the L1 norm, $\alpha \in [0,1]$ is the desired sparsity level, and $\textrm{diag}(\cdot)$ denotes a diagonal matrix obtained by setting the diagonal to the argument vector.
The loss function of Equation~\eqref{eq:LOUPE1} quantifies the average quality of the reconstruction.
The constraint ${\color{black}\frac{1}{d}\|\sigma_t(\bm{O})\|_1} = \frac{1}{d}\|\bm{P}\|_1 = \alpha$ ensures that the probabilistic under-sampling mask has an average value of $\alpha$, which will correspond to an acceleration factor of $R= 1/\alpha$.
We emphasize that the problem formulation of Equation~\eqref{eq:LOUPE1} is slightly different than the formulation in our conference paper~\cite{bahadir2019learning}, where we employed a sparsity inducing regularization instead of a hard constraint.
That formulation required fine-tuning a hyper-parameter in order to obtain the desired sparsity level, which was computationally inefficient.
In this paper we implement a practical strategy, described below, to solve the problem with the hard sparsity constraint directly.

The loss function involves an expectation over the random binary mask $\bm{M}$.
We approximate this expectation using a Monte Carlo-based sample averaging strategy:
{\color{black}
\begin{multline}
{
 \min_{\bm{O}, \theta} 
\sum_{j=1}^n \frac{1}{K}
\sum_{k=1}^K
\| A_\theta (F^{H} \textrm{diag}(\bm{m}^{(k)}) F \bm{x}_j) - \bm{x}_j \|_2^2},\\
{\textrm{such that } \frac{1}{d}\|\sigma_t(\bm{O})\|_1 = \alpha}
\label{eq:LOUPE2}\end{multline}}
where $\bm{m}^{(k)}$ are independent realizations drawn from $\prod_{i} \mathcal{B}(\sigma_t(\bm{O}_i))$.
Similar to the re-parameterization trick used in variational techniques, such as the VAE~\cite{kingma2013auto}, we can re-write Equation~\eqref{eq:LOUPE2}:
{\color{black}
\begin{multline}
\min_{\bm{O}, \theta}
\sum_{j=1}^n \frac{1}{K} 
\sum_{k=1}^K \| A_\theta (F^{H} \textrm{diag}(\bm{U}^{(k)} \leq \sigma_t(\bm{O})) F \bm{x}_j) - \bm{x}_j \|_2^2, 
\\
{\textrm{such that } \frac{1}{d}\|\sigma_t(\bm{O})\|_1 = \alpha}
\label{eq:LOUPE3}\end{multline}}
where $\bm{U}^{(k)}$ are independent realizations drawn from $\prod_{i=1}^d \mathcal{U}(0,1)$, a spatially independent set of uniform random variables on $[0,1]$. The result of the inequality operation is 1 if the condition is satisfied and 0 otherwise.
In Equation~\eqref{eq:LOUPE3} the random draws are thus from a constant distribution (independent uniform) and the probabilistic mask $\bm{P}$ {\color{black}(through its parameterization $\bm{O}$),} only effects the thresholding operation. 

\subsection{Implementation}
\begin{figure*}[t!]
\begin{centering}
\includegraphics[width=1.0\textwidth]{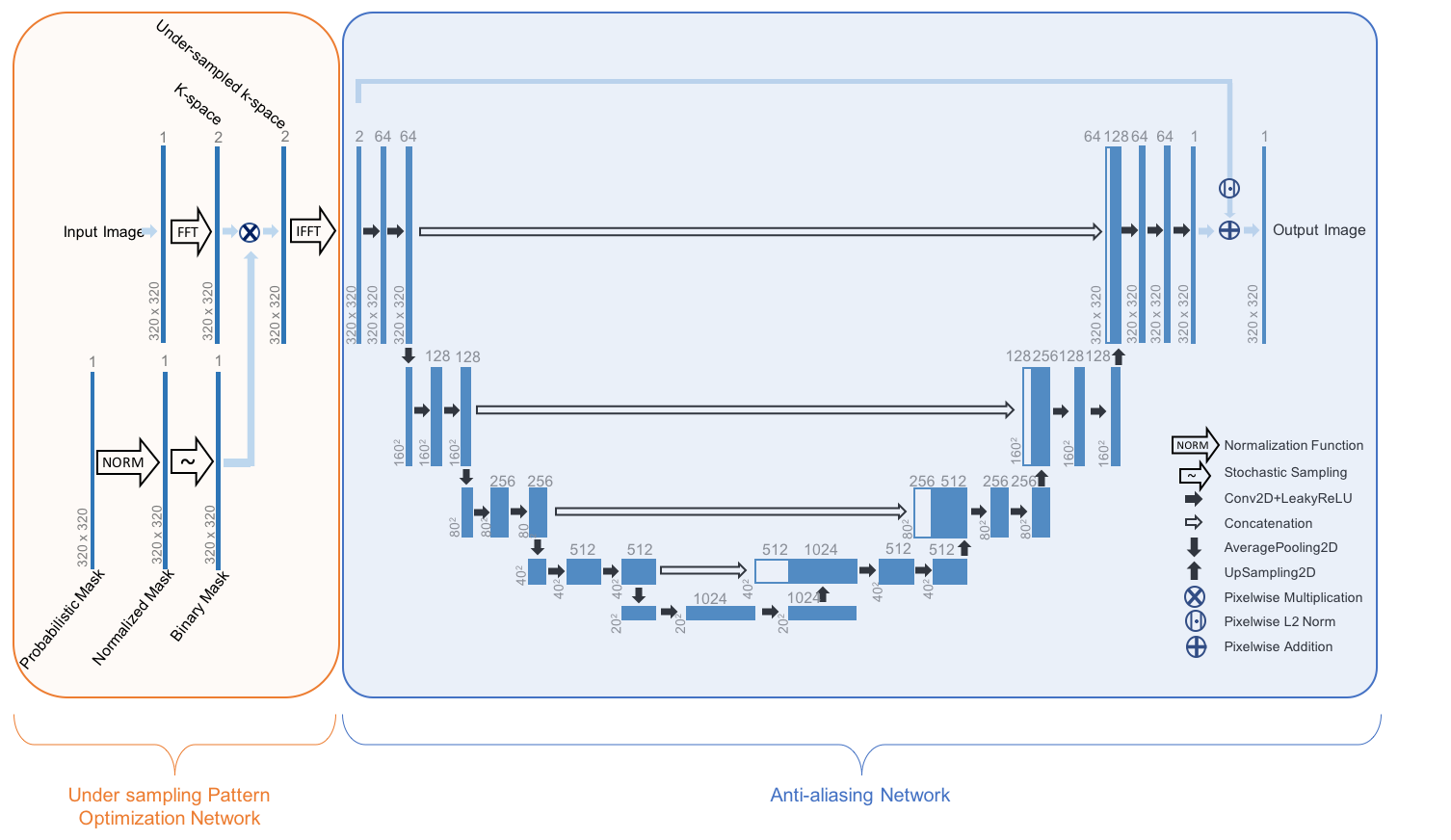}
\caption{LOUPE architecture with two building blocks: The Under-sampling Pattern Optimization Network and the Anti-aliasing Network. 
The model implements an end-to-end learning framework for optimizing the under-sampling pattern while learning the parameters for faithful reconstructions. 
The normalization layer for the probabilistic mask computes Equation~\eqref{eq:norm}.
Red arrows denote 2D convolutional layers with a kernel size of $3 \times 3$, followed by Leaky ReLU activation and Batch Normalization. 
Green vertical arrows represent average pooling operations and yellow vertical arrows refer to up-sampling. 
The initial number of channels in the U-Net is $64$, which doubles after each average pooling layer. 
The U-Net also uses skip connections depicted with gray horizontal arrows, that concatenate the corresponding layers to facilitate better information flow through the network. 
} 
\label{fig:u-net}
\end{centering}
\end{figure*}

In our first version of LOUPE, we used a 2D U-Net architecture~\cite{ronneberger2015u} for the anti-aliasing function, as depicted in Figure~\ref{fig:u-net} and similar to other prior work~\cite{lee2017deep,hyun2018deep}. 
To enable a differentiable loss function, we relaxed the non-differentiable threshold operation of Equation~\eqref{eq:LOUPE3} with a sigmoid. This relaxation is similar to the one used in recent Gumbel-softmax~\cite{jang2016categorical} and concrete distributions~\cite{maddison2016concrete} in training neural networks with discrete representations.
Our final objective is:
{\color{black}
\begin{multline}
\min_{\bm{O}, \theta}
\sum_{j=1}^n \frac{1}{K}
\sum_{k=1}^K
\| A_{\theta} (F^{H} \textrm{diag}(\sigma_s(\sigma_t(\bm{O})-\bm{U}^{(k)})) F \bm{x}_j) - \bm{x}_j \|_2^2,
\\
{\textrm{such that } \frac{1}{d}\|\sigma_t(\bm{O})\|_1 = \alpha}
\label{eq:LOUPE4}\end{multline}
}
where $\sigma_s(a) = \frac{1}{1+e^{-sa}}$ is a sigmoid with slope $s$.

A critical issue for LOUPE is the enforcement of the hard constraint ${\color{black} \frac{1}{d}\|\sigma_t(\bm{O})\|_1=}\frac{1}{d}\|\bm{P}\|_1 = \alpha$.
To achieve this we introduce a normalization layer, which rescales $\bm{P}$  to satisfy the constraint. 
We first define $\bar{p} = \frac{\|{\bm{P}}\|_1}{d}$, which is the average value of the pre-normalization probabilistic mask. 
Note that $1-\bar{p}$ is the average value of $1-\|{\bm{P}}\|_1$.
We define the normalization layer as:
{\color{black}
\begin{equation}
	N_{\alpha}(\bm{P})= 
    \begin{cases}
   \frac{\alpha}{\bar{p}}  {\bm{P}}, \textrm{ if } \bar{p} \geq \alpha \\
   1 - \frac{1-\alpha}{1-\bar{p}} ({\color{black}\bm{1}} - {\bm{P}}),  \textrm{ otherwise.}
    \end{cases}
    \label{eq:norm}
\end{equation}}
It can be shown that Equation~\eqref{eq:norm} yields $N_{\alpha}(\bm{P})\in [0,1]^d$ and $\frac{\|N_{\alpha}(\bm{P})\|_1}{d} = \alpha$.
This normalization trick allows us to convert the constrained problem of Equation~\eqref{eq:LOUPE4} to the following unconstrained problem:
\begin{multline}
\min_{\bm{O}, \theta}
\sum_{j=1}^n \frac{1}{K}
\sum_{k=1}^K 
\\
\| A_{\theta} (F^{H} \textrm{diag}(\sigma_s(N_{\alpha}(\sigma_t(\bm{O}))-\bm{U}^{(k)})) F \bm{x}_j) - \bm{x}_j \|_2^2.
\label{eq:LOUPE5}
\end{multline}

Figure~\ref{fig:u-net} illustrates an instantiation of the LOUPE model. 
{\color{black}In the depicted version of LOUPE, we used a U-Net architecture to implement $A_\theta$, the reconstruction network. 
We have also experimented with an alternative reconstruction model, namely Cascade-Net proposed in~\cite{schlemper2017deep}.
In the supplementary material, we present LOUPE results obtained with this alternative implementation. 
}
The LOUPE network accepts complex 2D images represented as two channels, corresponding to the imaginary and real components. 
The network applies a forward discrete Fourier transform, $F$, converting the data into k-space. 
The k-space measurements are under-sampled by $\tilde{\bm{m}} = \sigma_s(\bm{P}-\bm{u}^{(k)})$, which approximates a Monte Carlo realization of the probabilistic mask $\bm{P}$.
The values of the probabilistic mask are computed using the sigmoid operation of pixel-wise parameters $\bm{O}$ that are learned during the training process. 
The mask~$\tilde{\bm{m}}$ is multiplied element-wise with the k-space data (approximating retrospective under-sampling by inserting zeros at missing points), which is then passed to an inverse discrete Fourier transform $F^{H}$.

This image, which will typically contain aliasing artifacts, is then provided as input to the anti-aliasing network $A_\theta(\cdot)$, {\color{black} also  called the reconstruction network}.
This network takes a complex-valued under-sampled image, represented as two-channels, and aims to minimize the reconstruction error, such as the squared difference between the magnitude images of ground truth and the reconstruction. 
We can view the entire pipeline to be made up of two building blocks: the first optimizing the under-sampling pattern, and the second solving the reconstruction problem.


The model was implemented in Keras~\cite{chollet2015keras}, with Tensorflow~\cite{abadi2016tensorflow} in the back-end. 
Custom functions were adapted from the Neuron library~\cite{dalca2018anatomical}. ADAM~\cite{kingma2014adam} with an initial learning rate of $0.001$ was used for optimization, and learning was terminated when validation loss plateaued.
Consistent with prior work~\cite{kingma2013auto,jang2016categorical,maddison2016concrete}, a single Monte Carlo realization was drawn for each training datapoint. 
This is common practice in variational neural networks as it is a computationally efficient approach that yields an unbiased estimate of the gradient that is used in stochastic gradient descent. 
We used a batch size of {\color{black}16.}
{\color{black} We used a slope value of $s=200$ for $\sigma_s$ that approximates the thresholding operation and a slope value of $t=5$ for $\sigma_t$ that squashes the values of $\bm{O}$ to the range $[0, 1]$}. 
{\color{black} These values for the slope hyper-parameters were chosen based on a grid search strategy, where we identified the pair of values that yields the smallest validation loss (see Supplementary Material).
We note, however, that we did not observe a significant amount of sensitivity to these values.}
Our code is freely available at: \url{https://github.com/cagladbahadir/LOUPE/}

\section{Empirical Analysis} 
\label{sec:experiments}
\subsection{Data}

We conducted our experiments using the NYU fastMRI dataset~\cite{zbontar2018fastmri} (\underline
{fastmri.med.nyu.edu}). 
This is a freely available, large-scale, public data set of knee MRI scans.
The dataset originally comprises of 2D coronal knee scans acquired with two pulse sequences that result with Proton Density (PD) and Proton Density Fat Supressed (PDFS) weighted images. 
In our experiments, we used the provided emulated single-coil (ESC) k-space data from the Biograph mMR (3T) scanner, which were derived from raw 15-channel multi-coil data. 
We used 100 volumes from the provided official training dataset as our training dataset, and split the provided validation data into two halves, where we used 10 volumes for validation and 10 volumes for test.
We could not rely on the provided test data for evaluation as they were not fully sampled. 
For the scans we analyzed, the sequence parameters were: echo train length of 4, matrix size of 320x320, in plane resolution of $0.5 mm \times 0.5 mm$, slice thickness of $3 mm$ and no gap between slices. The time of repetition (TR) varied between $2200$ and $3000 ms$ and the Echo Time (TE) ranged between $27$ and $34 ms$. Training volumes had $38 \pm 4$ slices, where the validation volumes had $37\pm 3$ and test volumes had $38\pm 4$. 

Each set (training, validation and test) had differing slice sizes across volumes. 
After taking the Inverse Fourier Transform of the ESC k-space data, and rotating and flipping the images to match the orientation in the fastMRI paper~\cite{zbontar2018fastmri}, we cropped the central 320x320 and normalized by dividing to the maximum magnitude within each volume.

The U-Net architecture we employed was also used in the original NYU fastMRI study~\cite{zbontar2018fastmri} for reconstruction, with the only difference that their implementation accepted a single-channel, real-valued input image. 

\subsection{Evaluation}

We used three metrics in the quantitative evaluations of the reconstructions with respect to the ground truth images created from fully sampled measurements: (1) peak signal to noise ratio (PSNR), (2) structural similarity index (SSIM), and (3) high frequency error norm (HFEN).

PSNR is a widely used metric for evaluating the quality of reconstructions in compressed sensing applications~\cite{welstead1999fractal,sun2016deep}, 
defined as:
\begin{equation}
    {\rm PSNR}(\bm{x}, \hat{\bm{x}}) = 10 \log_{10}\frac{\max(\bm{x})^2 d}{\|\bm{x} - \hat{\bm{x}}\|_2^2},
\end{equation}
where $d$ is, as above, the total number of pixels in the full-resolution grid, and $\hat{\bm{x}}$ denotes the reconstruction of the ground truth full-resolution image $\bm{x}$.

SSIM aims to quantify the perceived image quality~\cite{wang2004image}:
\begin{equation}
    {\rm SSIM}(\bm{x},\hat{\bm{x}})=\frac{(2\mu_{\bm{x}}\mu_{\hat{\bm{x}}}+c_{1})+(2\sigma_{\bm{x}\hat{\bm{x}}}+c_{2})}{(\mu_{\bm{x}}^2+\mu_{\hat{\bm{x}}}^2+c_{1}) (\sigma_{\bm{x}}^2+\sigma_{\hat{\bm{x}}}^2+c_{2})},
\end{equation}
where $\mu_{\bm{x}}$ and $\mu_{\hat{\bm{x}}}$ are defined as the ``local'' average values for the original and reconstruction images, respectively,
computed within an $N \times N$ neighborhood.
Similarly, $\sigma_{\bm{x}}^2$ and $\sigma_{\hat{\bm{x}}}^2$ are the local variances; and $\sigma_{\bm{x}\hat{\bm{x}}}$ is local covariance between the reconstruction and ground truth images.
$c_{1}=(k_{1}L)^2$ and $c_{2}=(k_{2}L)^2$ are  constants that numerically stabilize the division, where $L$ is the dynamic range of the pixel values and $k_1$ and $k_2$ are user defined.
We use the parameters provided in~\cite{zbontar2018fastmri} for computing SSIM values: $k_{1}=0.01$, $k_{2}=0.03$, and a window size of $7\times7$.
The dynamic range of pixel values was computed over the entire volume.
%

Finally, we also report the high-frequency error norm (HFEN), which is used to quantify the quality of reconstruction of edges and fine features.
Following ~\cite{ravishankar2011mr}, we use a $15 \times 15$ Laplacian of Gaussian (LoG) filter with a standard deviation of $1.5$ pixels.
The HFEN is then computed as the  L2 difference between LoG filtered ground truth and reconstruction images.

\subsection{Benchmarks} 
We compared the LOUPE-optimized mask and other mask configurations, together with one machine learning (ML) based and three non-ML based  reconstruction techniques. 

\subsubsection{Reconstruction Methods}

Each reconstruction method was optimized on a representative slice from the data-set to find the values of hyper-parameters that yielded the best quality reconstruction for each of the individual masks. In other words, hyper-parameter tuning was done separately for each mask.

The first benchmark reconstruction method is BM3D~\cite{eksioglu2016decoupled}, which is an iterative algorithm that alternates between de-noising and reconstructions steps, and was shown to yield faithful reconstructions in the under-sampled cases.\footnote{We used the code at: \url{http://web.itu.edu.tr/eksioglue/pubs/BM3D\_MRI.htm}.} 
A grid search on the parameters: $\sigma$, noise and $\lambda$ were conducted, over the range $3 - 3\times 10^2$ and $0-10^{2}$, respectively. 

The second reconstruction benchmark method, called P-LORAKS, is based on low-rank modeling of local k-space neighborhoods~\cite{haldar2014low,haldar2016p}.\footnote{ 
We used the code available at: \url{https://mr.usc.edu/download/loraks2/}.}
A grid search on the parameters: $\lambda$, VCC (Virtual Conjugate Coils) and rank was conducted. The regularization parameter $\lambda$ was searched between $10^{-2}$ and 1. The rank value which is related to the non-convex regularization penalty was searched between 1 to 45 and the VCC, which is a Boolean parameter was searched for both cases: 1 and 0. 

The third benchmark method is Total Generalized Variation (TGV) based reconstruction with Shearlet Transform~\cite{guo2014new}. The method regularizes image regions while keeping the edge information intact.\footnote{We used the code at \url{http://www.math.ucla.edu/~wotaoyin/papers/tgv\_shearlet.html}.}
We conducted a grid search for the parameters $\beta$, $\lambda$, $\alpha_{\bm{0}}$, $\alpha_{\bm{1}}$ and  $\mu_{\bm{1}}$, adopting a range for each parameter based on the suggestions from the original paper~\cite{guo2014new}.
$\lambda$ was searched between $10^{-3}$ and $5 \times 10^{-1}$, $\beta$ was searched between $10^{2}$ and $10^{3}$. The parameters $\mu_{\bm{1}}$, $\alpha_{\bm{0}}$ and $\alpha_{\bm{1}}$ were respectively searched between values: $3\times10^{2}$ and $10 \times {3}$, $8 \times 10^{-4}$ to $10^{-2}$ and $10^{-3}$ to $10^{-1}$.

The final benchmark reconstruction method is the residual U-Net~\cite{lee2017deep}, which is also a building block in our model. 
In this framework, the U-Net is used as an anti-aliasing neural network widely used for biomedical image applications. 
Unlike in LOUPE, the benchmark U-Net implementation is trained for a fixed under-sampling mask, that is provided by the user. 
In our experiments the U-Net models were individually trained and tested for each mask configuration.

\subsubsection{Under-sampling Masks}

\begin{figure*}[t]
\begin{center}
\includegraphics[width=1.5\columnwidth]{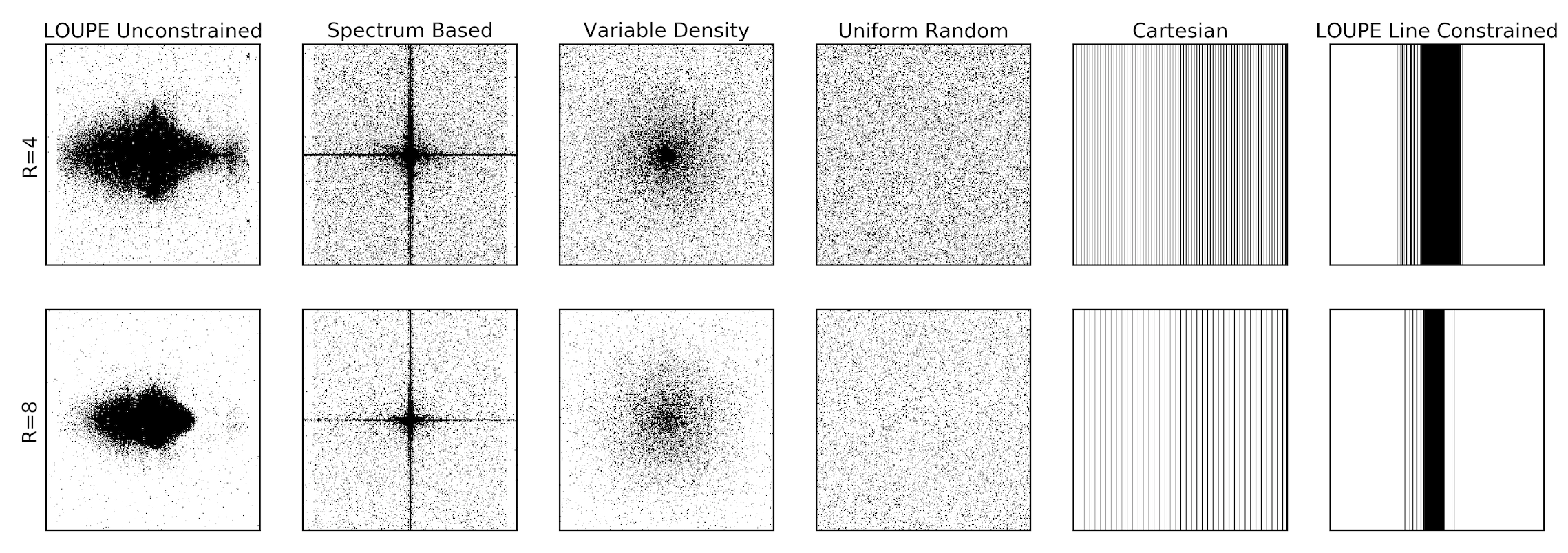}
\caption{{\color{black}LOUPE-optimized and benchmark masks for two levels of acceleration rates for NYU fastMRI data set: $R=4$ and $R=8$. Black dots represent k-space points that are sampled and white areas correspond to measurements that are not acquired. {\color{black}The two right-most masks represent 1D under-sampling strategies, whereas the remaining four represent 2D under-sampling. These two categories of under-sampling rely on different acquisition protocols.}}}
\label{fig:maskknee}
\end{center}
\end{figure*}

We implemented several widely used benchmark under-sampling masks, shown in Figure~\ref{fig:maskknee}. 
{\color{black}The first set of masks we considered are what we refer to as 2D Cartesian under-sampling masks.
These are physically feasible for 2D imaging with (i) two phase-encoding dimensions and no frequency-encoding gradient; or (ii) for 3D acquisition with one frequency encoding dimension and two phase-encoding dimensions.
In this category, we have ``Random Uniform''~\cite{gamper2008compressed}, }which exhibits the benefits of stochastic under-sampling in creating incoherent, noise-like artifacts~\cite{lustig2008compressed}, thus facilitating the discrimination of artifacts from the signal present in the image. 
We also employed a parametric Variable Density~\cite{wang2010variable} (VD) mask that samples lower frequencies more densely, while still enjoying the benefits of creating noise-like artifacts.
{\color{black}We used the publicly available code of~\cite{lustig2007sparse} to identify the optimal values of the VD mask that maximizes incoherence (of aliasing artifacts) in the wavelet domain.
Another 2D under-sampling mask we implemented was a data-driven strategy that was based on~\cite{vellagoundar2015robust}. 
In our implementation, we averaged the magnitude spectrum of all the fully-sampled training data.
The average spectrum was then thresholded to keep the k-space points that have the largest average magnitude. 
We refer to this mask as ``spectrum-based.''
}

{\color{black}Next, we considered 1D under-sampling masks, which are physically feasible for 2D imaging with one frequency-encoding dimension and one phase-encoding dimension.}
Cartesian~\cite{haldar2011compressed} with skipped lines {\color{black}orthogonal to the provided read-out direction} is a popular mask in this category.
{\color{black}In addition, we implemented a version of LOUPE where the mask was constrained to be made up of lines along the read-out direction, which is similar to a strategy proposed in~\cite{weiss2019learning}.
We achieved this with a simple modification in our code by sharing the weights corresponding to the entries of $\bm{O}$ along each read-out line.
This ``line-constrained version'' of LOUPE, we believe, is one step closer to being physically realistic.
When we need to be explicit, we refer to the first LOUPE version as unconstrained.}

{\color{black}
We underscore that it might not be fair to compare the 1D and 2D under-sampling masks, as the two categories would rely on different acquisition protocols.
Thus, any comparisons we present below need to account for this difference.}



\subsection{Results}

{\color{black}Figure~\ref{fig:maskknee} shows the LOUPE optimized and benchmark masks}, for two different acceleration rates $R=4$ and $R=8$.
%
The LOUPE, {\color{black} spectrum-based}, and variable density masks share the behavior of a drop in density from lower to higher frequencies, however differ significantly in their symmetry and shape. 
Although the unconstrained LOUPE-optimized masks are similar to the VD masks in terms of emphasizing lower frequencies, importantly lateral frequencies are favored significantly more than ventral/dorsal frequencies. 
{\color{black}The data-driven spectrum-based masks, on the other hand, exhibit a similar tendency to pick out more lateral frequencies, yet with a dramatically different overall shape that concentrates around the two axes.
The line-constrained LOUPE masks also show a strong preference for lower frequencies.
}

\begin{figure}[t]
\begin{center}
\includegraphics[width=1.0\columnwidth]{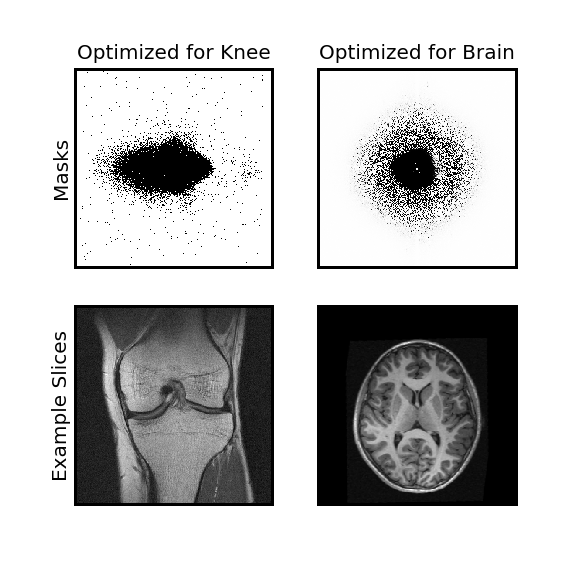}
\caption{LOUPE-optimized under-sampling masks (for $R=8$) compared side by side for the knee and brain anatomies. 
The brain mask was derived using the data described in our prior work~\cite{bahadir2019learning}. 
There is a striking difference in the symmetry, which can be appreciated visually. Note that black dots represent k-space points that are sampled and white areas correspond to measurements that are not acquired.}
\label{fig:maskcomp}
\end{center}
\end{figure}

In an earlier conference paper~\cite{bahadir2019learning}, we had reported an unconstrained LOUPE-optimized mask for T1-weighted brain MRI scans, obtained from a public dataset~\cite{di2014autism}.
Figure~\ref{fig:maskcomp} shows a side by side comparison of the two LOUPE-optimized masks for the knee and brain anatomies, respectively, emphasizing the asymmetry present in the knee dataset.  The knee mask favors lateral frequencies more than ventral/dorsal frequencies due to the unique features of knee anatomy, where there is significantly more tissue contrast in the lateral direction. Brain scans, on the other hand, exhibit a more radially symmetric behavior. This comparison highlights the importance of a data-driven approach in identifying the under-sampling mask.

\begin{figure*}[p]
\begin{centering}
\includegraphics[width=0.9\textwidth]{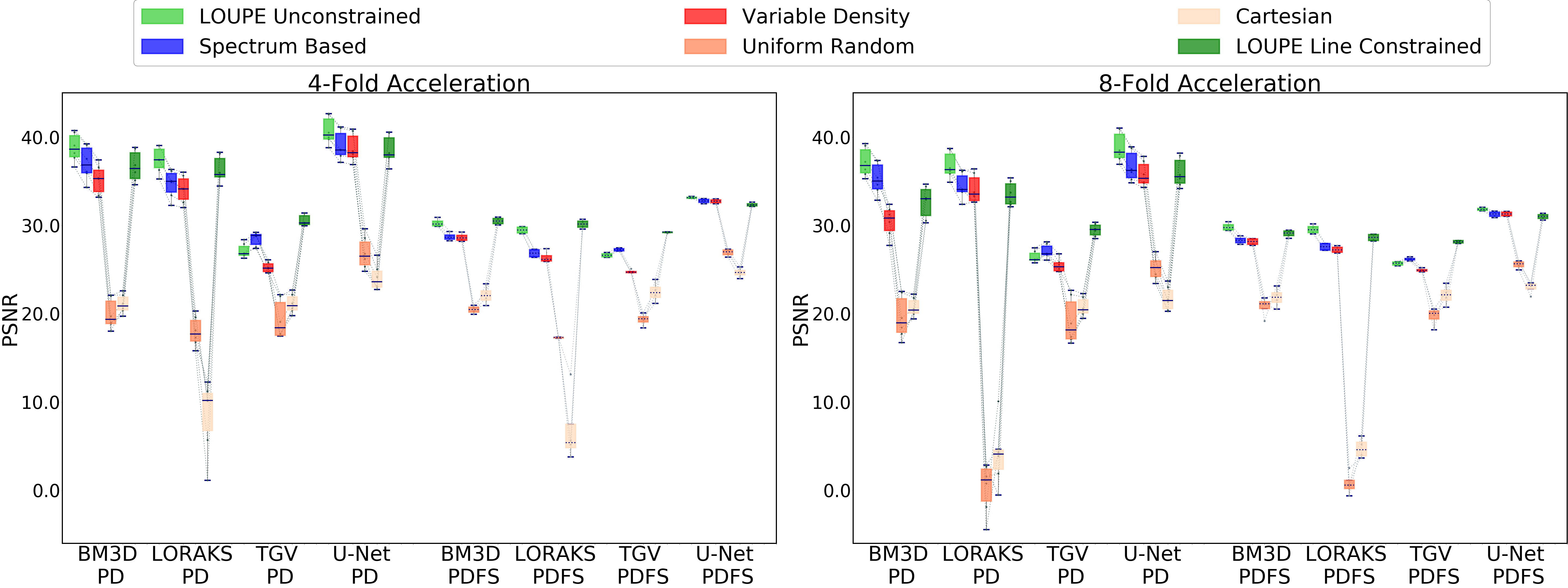}
\includegraphics[width=0.9\textwidth]{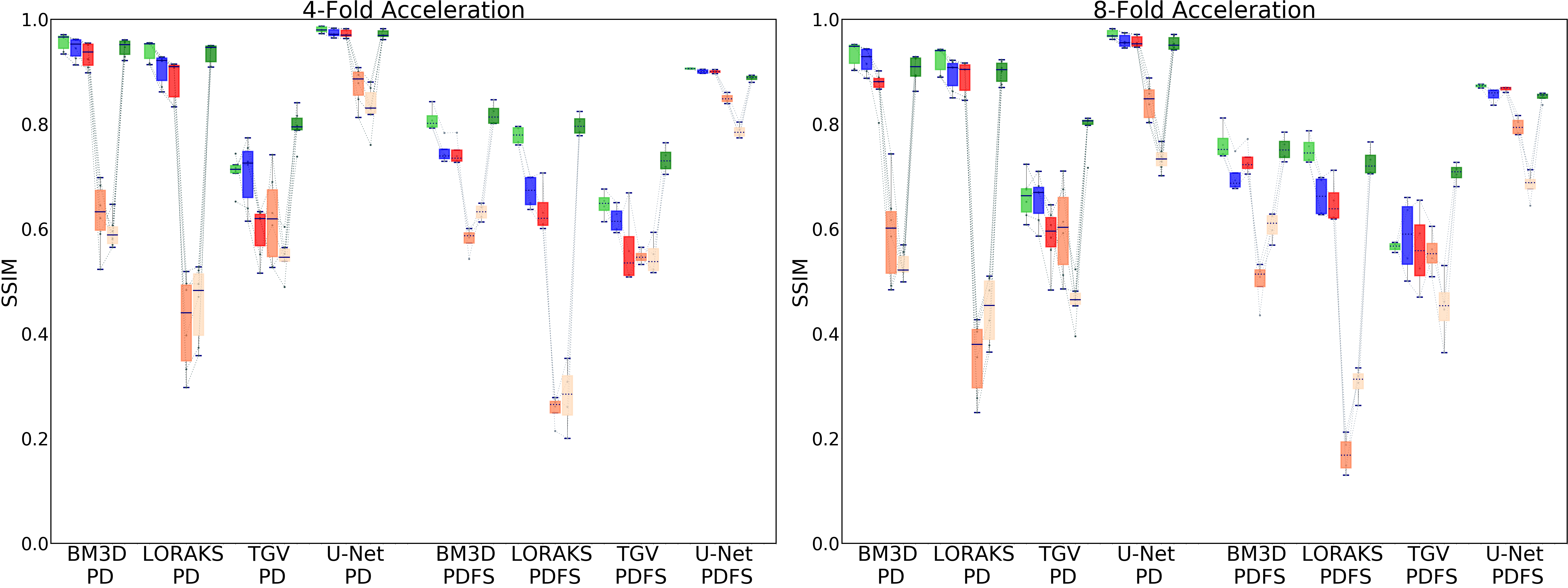}
\includegraphics[width=0.9\textwidth]{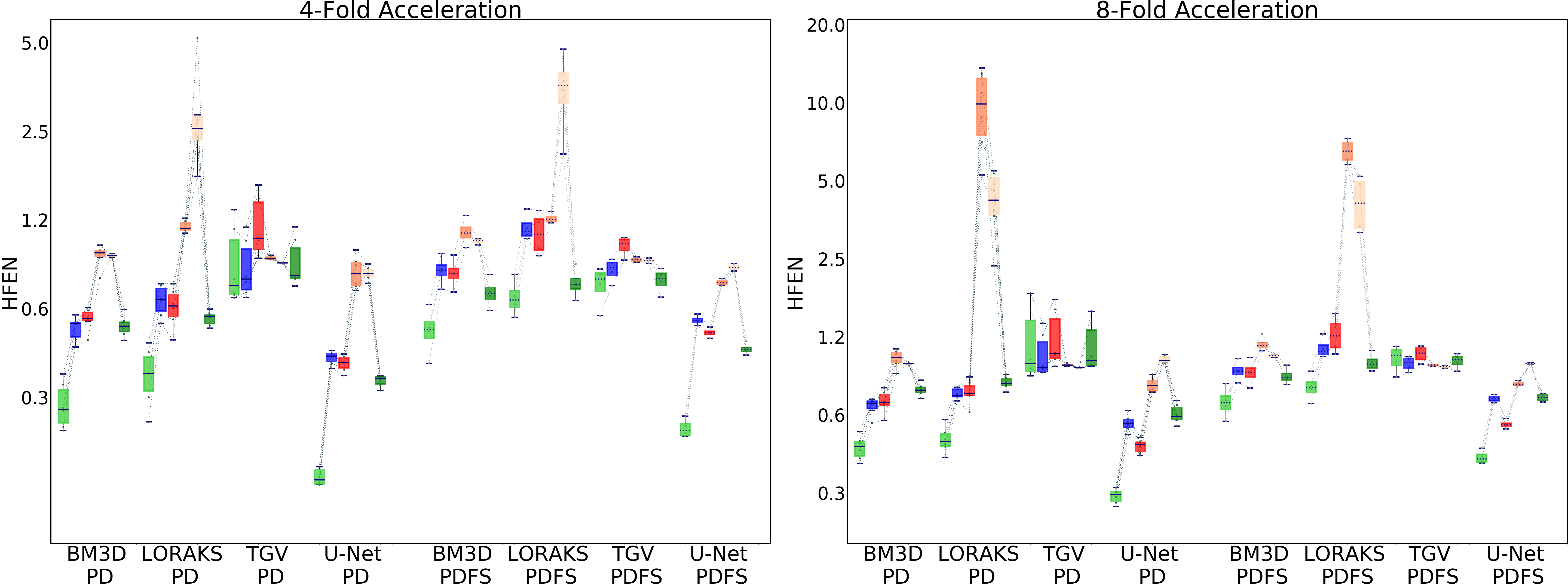}
\caption{
\textbf{Quantitative evaluation of reconstruction quality for the NYU fastMRI data set.} Top: PSNR values (higher is better), middle: SSIM results (higher is better), bottom: high-frequency error norm (HFEN) results (lower is better).
For each plot, we show four reconstruction methods using different acquisition masks, including LOUPE-optimized masks in green. 
{\color{black}Note that 2D (first four in each group) and 1D (last two in each group) under-sampling masks should be interpreted separately and any cross-category comparison should be done with caution.}
The slice-averaged values for each test subject are connected with lines.
For each box, the blue straight line shows the median value.
Patients with Proton Density (PD) images and Proton Density Fat Suppressed (PDFS) images are shown separately. 
The whiskers indicate the the most extreme (non-outlier) data points.
}
\label{fig:quantitative-results}
\end{centering}
\end{figure*}

Figure~\ref{fig:quantitative-results} shows subject-level quantitative reconstruction quality values computed for the different masks and reconstruction methods, under two different acceleration conditions.
{\color{black}We present the results for the 6 test subjects with PD weighted images and 4 test subjects with PDFS weighted images, separately.} 
The fat suppression operation inherently lowers the signal level, as fat has the highest levels of signal in an MRI scan, thus yielding a noisier image, where small details are more apparent. 
In contrast, regular PD weighted scans that contain fat tissue have inherently higher SNR. 
Therefore, the PD scans yield better quality reconstructions compared to the PDFS scans. 

Figure~\ref{fig:quantitative-results} {\color{black}reveals} that overall LOUPE yields significantly better results than competing masks, when used with any of the tested reconstruction methods and under the two acceleration rates. 
Furthermore, the U-Net reconstruction model, coupled with LOUPE-optimized masks yields the best reconstructions, for both acceleration rates and for all test subjects.
Unsurprisingly, the unconstrained version of LOUPE is generally better than the line-constrained version, with the exception of the TGV reconstruction method, where the line-constrained LOUPE outperformed unconstrained LOUPE. 
%
These results suggest that LOUPE-optimized masks can offer a performance gain even when not used with the U-Net based reconstruction method.
We further observe that the Variable Density {\color{black} and spectrum-based masks} outperforms the uniform mask, which is consistent with prior literature. 


Table~\ref{table:table_perf} lists the summary quantitative values for each of the reconstruction method and mask configurations, where averages within the PD and PDFS subjects are presented separately.
Supplementary Tables 1 and 2 includes standard deviations for these results.
The lowest HFEN and highest PSNR and SSIM values are bolded for each reconstruction method, metric and subject group. 
We observe that the LOUPE-optimized masks yield the highest reconstruction quality under all considered conditions, and the line-constrained and unconstrained versions are often quite similar. 
For reference, we also added the best values reported in~\cite{zbontar2018fastmri} that was achieved with a U-Net reconstruction model and a Variable Density Cartesian mask. 
We underscore that their model was trained on the full training set and evaluated on the full validation set. 
{\color{black}Nevertheless, we observe that their results are consistent with ours, often falling between our U-Net models trained for Cartesian and VD masks, but always under-performing compared to the U-Net model trained with LOUPE masks.}

\begin{figure*}
\begin{centering}
\includegraphics[width=\textwidth]{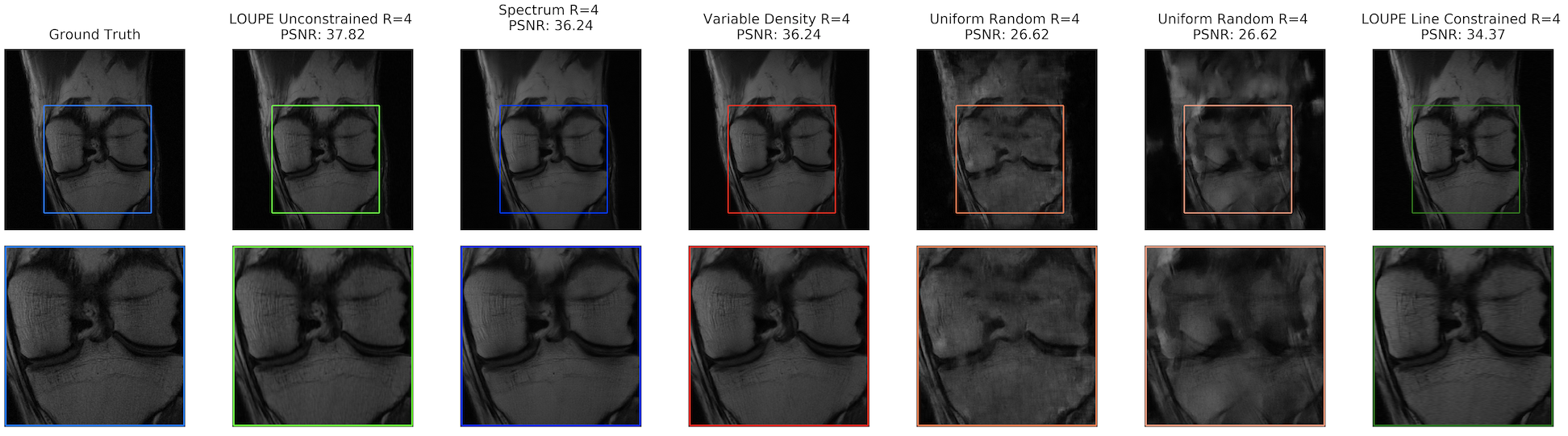}
\caption{U-Net based reconstructions for an example PD slice from NYU fastMRI experiments with 4-fold acceleration rate ($\alpha=0.25$). 
Each column corresponds to an under-sampling mask. 
{\color{black}Corresponding PSNR values are listed above each image.}}
\label{fig:Knee-4x}
\end{centering}
\end{figure*}

%

\begin{figure*}
\begin{centering}
\includegraphics[width=\textwidth]{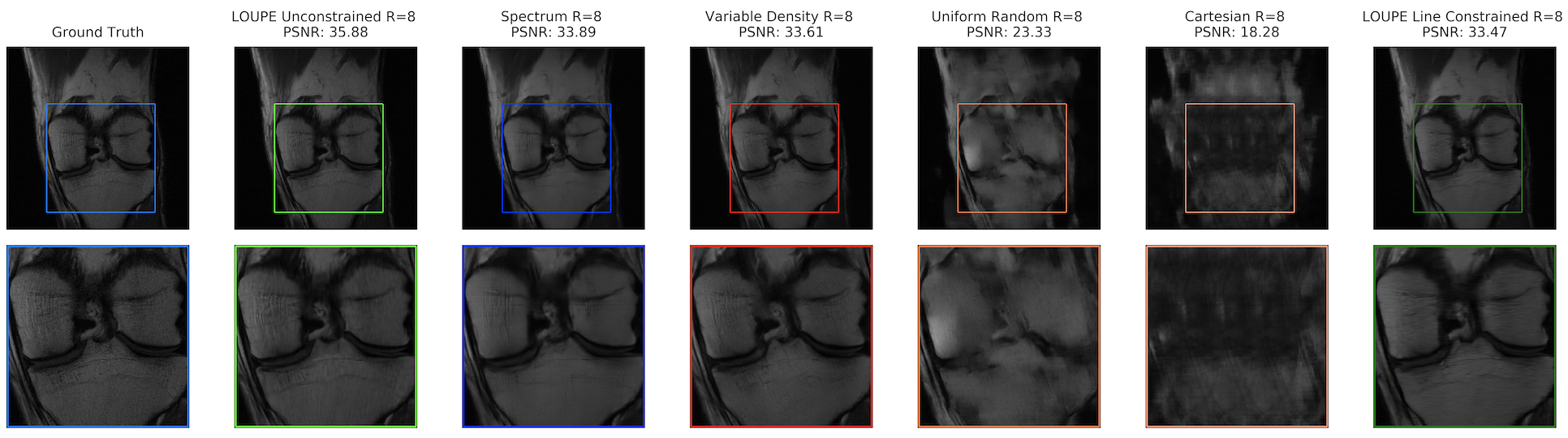}
\caption{U-Net based reconstructions for an example PD slice from NYU fastMRI experiments with 8-fold acceleration rate ($\alpha=0.125$). 
Each column corresponds to an under-sampling mask.
{\color{black}Corresponding PSNR values are listed above each image.}}
\label{fig:Knee-8x}
\end{centering}
\end{figure*}


Figures~\ref{fig:Knee-4x} and \ref{fig:Knee-8x} show example PD weighted slices and U-Net based reconstructions, obtained after under-sampling with $R=4$ and $R=8$, using different masks.
As can be appreciated from these figures, much of the structural details that correspond to the femur, tibia, tendons, ligaments, muscle tissue, and meniscus are more faithfully captured in the reconstructions obtained with the LOUPE-optimized masks, whereas reconstructions computed with other masks were more blurry.
 For example, 
 edges of the posterior cruciate ligament and lateral meniscus suffer from blurring artifacts in the benchmark mask configurations compared to the LOUPE-optimized masks. 
 Overall, in agreement with the quantitative results presented above, the LOUPE-optimized mask yields the best-looking reconstruction results and these reconstructions appear less prone to artifacts and blurring, while preserving more of the anatomical detail. 

\begin{sidewaystable*}[t]
\caption{{\color{black}Quantitative results of reconstruction quality, averaged over test cases. 1D masks are separated from 2D masks and comparisons should be done accordingly. MSE: Mean squared error. MAE: Mean absolute error. HFEN: High frequency error norm. PSRN: Peak signal to noise ratio. SSIM: Structural similarity metric.}} \label{table:table_perf} 
\centering \footnotesize
\begin{tabular}{lllllllllllll}
\textbf{Acceleration}    & Method                  & Masks                      & MSE PD           & MSE PDFS         & MAE PD          & MAE PDFS        & HFEN PD         & HFEN PDFS       & PSNR PD        & PSNR PDFS      & SSIM PD       & SSIM PDFS     \\ \hline
\multirow{25}{*}{4-fold} & \multirow{6}{*}{BM3D}   & Cartesian  (1D) Skipped Lines                & 0.00796          & 0.00629          & 0.0646          & 0.0587          & 0.9484          & 1.0596          & 21.1           & 22.1           & 0.59          & 0.63          \\
                         &                         & LOUPE Line Constrained     & 0.00023          & \textbf{0.00089} & 0.0105          & \textbf{0.0234} & 0.548           & 0.7096          & 36.69          & \textbf{30.5}  & 0.95          & \textbf{0.82} \\
                         \cline{3-13} 
                         &                         & Uniform Random             & 0.01074          & 0.00902          & 0.0707          & 0.0695          & 0.9467          & 1.1416          & 19.95          & 20.47          & 0.63          & 0.58          \\
                         &                         & Variable Density           & 0.00032          & 0.00137          & 0.0128          & 0.029           & 0.5772          & 0.8288          & 35.2           & 28.66          & 0.93          & 0.75          \\
                         &                         & Spectrum                   & 0.00021          & 0.00135          & 0.0107          & 0.0287          & 0.537           & 0.8464          & 37.05          & 28.72          & 0.94          & 0.75          \\
                         &                         & LOUPE Unconstrained        & \textbf{0.00014} & 0.00094          & \textbf{0.0088} & 0.0241          & \textbf{0.2972} & \textbf{0.5298} & \textbf{38.81} & 30.28          & \textbf{0.96} & 0.81          \\ \cline{2-13} 
                         & \multirow{6}{*}{LORAKS} & Cartesian   (1D) Skipped Lines               & 0.22717          & 0.26113          & 0.1785          & 0.1896          & 2.8759          & 3.5001          & 8.45           & 6.93           & 0.46          & 0.28          \\
                         &                         & LOUPE Line Constrained     & 0.00025          & \textbf{0.00097} & 0.0112          & \textbf{0.0246} & 0.5801          & 0.7668          & 36.32          & \textbf{30.14} & \textbf{0.94} & \textbf{0.8}  \\
                          \cline{3-13} 
                         &                         & Uniform Random             & 0.01692          & 0.01861          & 0.0913          & 0.1052          & 1.1886          & 1.2677          & 17.99          & 17.3           & 0.42          & 0.26          \\
                         &                         & Variable Density           & 0.00041          & 0.00231          & 0.0152          & 0.0372          & 0.6364          & 1.1334          & 34.1           & 26.4           & 0.89          & 0.64          \\
                         &                         & Spectrum                   & 0.00036          & 0.00208          & 0.0142          & 0.0355          & 0.6691          & 1.184           & 34.69          & 26.85          & 0.91          & 0.67          \\
                         &                         & LOUPE Unconstrained        & \textbf{0.00019} & 0.00113          & \textbf{0.0105} & 0.0266          & \textbf{0.3754} & \textbf{0.6854} & \textbf{37.42} & 29.48          & \textbf{0.94} & 0.78          \\ \cline{2-13} 
                         & \multirow{6}{*}{TV}     & Cartesian   (1D) Skipped Lines               & 0.00789          & 0.00582          & 0.0643          & 0.0564          & 0.8938          & 0.9126          & 21.14          & 22.46          & 0.55          & 0.55          \\
                         &                         & LOUPE Line Constrained     & \textbf{0.00089} & \textbf{0.0012}  & \textbf{0.0217} & \textbf{0.027}  & 0.9029          & 0.7823          & \textbf{30.57} & \textbf{29.22} & \textbf{0.8}  & \textbf{0.73} \\
                          \cline{3-13} 
                         &                         & Uniform Random             & 0.01286          & 0.01178          & 0.0755          & 0.0818          & 0.9303          & 0.9184          & 19.33          & 19.33          & 0.62          & 0.55          \\
                         &                         & Variable Density           & 0.003            & 0.00333          & 0.0417          & 0.0454          & 1.2104          & 1.0221          & 25.26          & 24.78          & 0.6           & 0.56          \\
                         &                         & Spectrum                   & 0.00144          & 0.00188          & 0.0289          & 0.0344          & 0.8695          & 0.8492          & 28.48          & 27.27          & 0.71          & 0.62          \\
                         &                         & LOUPE Unconstrained        & 0.00196          & 0.00218          & 0.0332          & 0.0366          & \textbf{0.8991} & \textbf{0.7559} & 27.13          & 26.63          & 0.71          & 0.65          \\ \cline{2-13} 
                         & \multirow{6}{*}{U-Net}  & Cartesian   (1D) Skipped Lines               & 0.00406          & 0.00345          & 0.0381          & 0.0401          & 0.8265          & 0.8648          & 24.11          & 24.65          & 0.83          & 0.79          \\
                         &                         & LOUPE Line Constrained     & 0.00015          & 0.00058          & 0.0084          & 0.0186          & 0.3563          & 0.4567          & 38.53          & 32.36          & 0.97          & 0.89          \\
                          \cline{3-13} 
                         &                         & Uniform Random             & 0.0022           & 0.00204          & 0.0303          & 0.0323          & 0.8342          & 0.7689          & 26.89          & 26.93          & 0.87          & 0.85          \\
                         &                         & Variable Density           & 0.00014          & 0.00053          & 0.0086          & 0.018           & 0.4088          & 0.5186          & 38.79          & 32.74          & 0.97          & 0.9           \\
                         &                         & Spectrum                   & 0.00013          & 0.00053          & 0.0084          & 0.0179          & 0.4266          & 0.5727          & 39.05          & 32.77          & 0.97          & 0.9           \\
                         &                         & LOUPE Unconstrained        & \textbf{0.00009} & \textbf{0.00048} & \textbf{0.0071} & \textbf{0.0173} & \textbf{0.1681} & \textbf{0.246}  & \textbf{40.71} & \textbf{33.16} & \textbf{0.98} & \textbf{0.91} \\ \cline{2-13} 
                         & U-Net NYU               & Variable Density Cartesian &                  &                  &                 &                 &                 &                 & 34.01          & 29.95          & 0.82          & 0.64          \\ \hline
\multirow{25}{*}{8-fold} & \multirow{6}{*}{BM3D}   & Cartesian   (1D) Skipped Lines               & 0.00871          & 0.00671          & 0.0665          & 0.0603          & 0.9855          & 1.0604          & 20.71          & 21.84          & 0.53          & 0.6           \\
                         &                         & LOUPE Line Constrained     & 0.00059          & 0.00124          & 0.0152          & 0.0273          & 0.7843          & 0.8844          & 32.66          & 29.08          & 0.9           & 0.75          \\
                          \cline{3-13} 
                         &                         & Uniform Random             & 0.01248          & 0.00854          & 0.0759          & 0.0683          & 1.0331          & 1.1762          & 19.56          & 20.81          & 0.59          & 0.5           \\
                         &                         & Variable Density           & 0.00097          & 0.00153          & 0.0217          & 0.0306          & 0.706           & 0.9147          & 30.45          & 28.16          & 0.87          & 0.73          \\
                         &                         & Spectrum                   & 0.00032          & 0.00148          & 0.0128          & 0.0301          & 0.6758          & 0.9278          & 35.26          & 28.32          & 0.92          & 0.7           \\
                         &                         & LOUPE Unconstrained        & \textbf{0.0002}  & \textbf{0.00104} & \textbf{0.0103} & \textbf{0.0256} & \textbf{0.4684} & \textbf{0.7022} & \textbf{37.16} & \textbf{29.85} & \textbf{0.93} & \textbf{0.76} \\ \cline{2-13} 
                         & \multirow{6}{*}{LORAKS} & Cartesian    (1D) Skipped Lines              & 0.4987           & 0.34232          & 0.2875          & 0.2739          & 4.2092          & 4.1481          & 4.05           & 4.77           & 0.44          & 0.31          \\
                         &                         & LOUPE Line Constrained     & 0.00046          & 0.00138          & 0.0147          & 0.029           & 0.8339          & 0.9979          & 33.56          & 28.63          & 0.9           & 0.73          \\
                          \cline{3-13} 
                         &                         & Uniform Random             & 1.15118          & 0.86483          & 0.3866          & 0.3221          & 9.7904          & 6.5299          & 0.26           & 0.77           & 0.35          & 0.17          \\
                         &                         & Variable Density           & 0.00041          & 0.00188          & 0.0152          & 0.0343          & 0.7671          & 1.2851          & 34.12          & 27.27          & 0.89          & 0.65          \\
                         &                         & Spectrum                   & 0.00038          & 0.00175          & 0.0145          & 0.033           & 0.7555          & 1.1352          & 34.45          & 27.59          & 0.9           & 0.66          \\
                         &                         & LOUPE Unconstrained        & \textbf{0.00022} & \textbf{0.00111} & \textbf{0.0111} & \textbf{0.0266} & \textbf{0.5038} & \textbf{0.8037} & \textbf{36.79} & \textbf{29.56} & \textbf{0.92} & \textbf{0.75} \\ \cline{2-13} 
                         & \multirow{6}{*}{TV}     & Cartesian     (1D) Skipped Lines             & 0.0086           & 0.00631          & 0.0664          & 0.0585          & \textbf{0.9508} & \textbf{0.9606} & 20.76          & 22.11          & 0.46          & 0.45          \\
                         &                         & LOUPE Line Constrained     & \textbf{0.00114} & \textbf{0.00153} & \textbf{0.0245} & \textbf{0.0302} & 1.1597          & 1.0106          & \textbf{29.48} & \textbf{28.15} & \textbf{0.79} & \textbf{0.71} \\
                          \cline{3-13} 
                         &                         & Uniform Random             & 0.01392          & 0.01097          & 0.0792          & 0.0785          & 0.975           & 0.9729          & 19.15          & 19.7           & 0.6           & 0.55          \\
                         &                         & Variable Density           & 0.00286          & 0.0032           & 0.0414          & 0.0445          & 1.2487          & 1.076           & 25.48          & 24.95          & 0.58          & 0.56          \\
                         &                         & Spectrum                   & 0.00199          & 0.0024           & 0.034           & 0.0388          & 1.0703          & 0.9868          & 27.07          & 26.21          & 0.66          & 0.59          \\
                         &                         & LOUPE Unconstrained        & 0.00229          & 0.0027           & 0.0364          & 0.0407          & 1.2016          & 1.0351          & 26.44          & 25.69          & 0.66          & 0.57          \\ \cline{2-13} 
                         & \multirow{6}{*}{U-Net}  & Cartesian   (1D) Skipped Lines               & 0.007            & 0.00509          & 0.0517          & 0.0489          & 1.0244          & 0.988           & 21.72          & 22.97          & 0.73          & 0.68          \\
                         &                         & LOUPE Line Constrained     & 0.00027          & 0.00079          & 0.0108          & 0.0214          & 0.6327          & 0.7301          & 36.0           & 31.01          & 0.95          & 0.85          \\
                          \cline{3-13} 
                         &                         & Uniform Random             & 0.00316          & 0.00279          & 0.0363          & 0.0376          & 0.8189          & 0.8267          & 25.17          & 25.56          & 0.84          & 0.8           \\
                         &                         & Variable Density           & 0.00027          & 0.00074          & 0.0118          & 0.0211          & 0.4746          & 0.5759          & 35.82          & 31.32          & 0.96          & 0.87          \\
                         &                         & Spectrum                   & 0.00023          & 0.00075          & 0.0105          & 0.021           & 0.5828          & 0.7233          & 36.69          & 31.28          & 0.96          & 0.86          \\
                         &                         & LOUPE Unconstrained        & \textbf{0.00014} & \textbf{0.00066} & \textbf{0.0085} & \textbf{0.0199} & \textbf{0.3044} & \textbf{0.4303} & \textbf{38.85} & \textbf{31.83} & \textbf{0.97} & \textbf{0.87} \\ \cline{2-13} 
                         & U-Net NYU               &                            &                  &                  &                 &                 &                 &                 & 31.5           & 28.71          & 0.76          & 0.56      
\end{tabular}
\end{sidewaystable*}


{\color{black}In the Supplementary Material, we present our results obtained with a version of LOUPE where the U-Net reconstruction model was replaced with Cascade-Net~\cite{schlemper2017deep} with the same architecture details of the reference paper.
Everything else in LOUPE, including all optimization settings were kept the same.
We observe that the optimized masks are very similar between U-Net-LOUPE and Cascade-Net-LOUPE. 
We also quantified the quality of reconstructions with these masks and found that there was no substantial difference. 
These results, we believe, demonstrate that LOUPE can be relatively robust to the choice of the reconstruction architecture. 
{\color{black}We note, however, that we have not experimented with more advanced reconstruction networks, such as the recently proposed~\cite{ramzi2020benchmarking}, which might yield different results.
We consider the exploration of different architectural designs for LOUPE as an important direction for future research.}
}

\section{Discussion}
\label{sec:conc}

Compressed sensing is a promising approach to accelerate MRI and make the technology more accessible and affordable. 
Since its introduction over a decade ago, it has gained in popularity and been used in a range of application domains.
Compressed sensing MRI has two fundamental challenges. 
The first is identifying where in k-space to sample (acquire) from. 
{\color{black}One approach to solve this problem is to take a data-driven strategy, where the optimal under-sampling pattern is identified using, for example, a collection of full-resolution scans that are retrospectively under-sampled.}
The second core problem of compressed sensing MRI is solving the ill-posed inverse problem of reconstructing a high quality image from under-sampled measurements.
This is conventionally solved via a regularized optimization approach, and more recently deep learning techniques have been proposed to achieve superior performance.

In this paper, we tackled both of these problems simultaneously.
We leveraged recently proposed deep learning based reconstruction techniques to design an end-to-end learning technique that enabled us to optimize the under-sampling pattern on given training data.
Our approach relies on full-resolution data that is under-sampled retrospectively, which is then fed to a reconstruction network.
The final output is then compared to the original input to assess overall quality.
We formulate an objective that optimizes reconstruction quality with a constrained number of collected measurements.
The stronger the constraint, the more aggressive acceleration rates we can achieve.

Our results indicate that LOUPE, the proposed method, can compute an optimized under-sampling pattern that is data-dependent and performs well with different reconstruction methods.
In this paper, we present empirical results obtained with a large-scale, publicly available knee MRI dataset.
Across all conditions we experimented with,  LOUPE-optimized masks coupled with the U-Net based reconstruction model offered the most superior reconstruction quality.
Even with an aggressive 8-fold acceleration rate, LOUPE's reconstructions contained much of the anatomical detail that was missed by alternative masks as seen in Figure~\ref{fig:Knee-8x}.

{\color{black}In our primary experiments, we used a U-Net based reconstruction method in the implementation of LOUPE. 
Additionally, we investigated the use of an alternative reconstruction network, namely Cascade-Net. 
Our results, which we present in the Supplementary Material, indicate that both versions of LOUPE achieved similar optimized masks and yielded similar reconstruction quality.
However, we consider the exploration of different architectural designs (such as ADMM-Net~\cite{sun2016deep} or MoDL~\cite{aggarwal2018modl}) for LOUPE as an important direction for future research.}

The LOUPE-optimized under-sampling mask for the analyzed knee MRI scans exhibited an interesting asymmetric pattern, which emphasized higher frequencies along the lateral direction much more than those along the vertical direction. 
This is likely due the anisotropy in the field of view of sagittal knee images.
This is in stark contrast to the LOUPE-optimized mask that we recently presented at a conference for T1-weighted brain MRI scans~\cite{bahadir2019learning}.
The brain mask exhibited a radially symmetric nature, very similar to a variable density mask, which is widely used in compressed sensing MRI.
This comparison highlights the importance of adapting the under-sampling pattern to the data and application at hand.

An important caveat of the presented LOUPE framework is that we largely ignored the costs of physically implementing the under-sampling pattern in an MR pulse sequence.
In {\color{black} the unconstrained version of LOUPE}, the cost of an under-sampling pattern is merely captured by the number of collected measurements.
We underscore that in 3D, the LOUPE-optimized mask can be implemented as a Cartesian trajectory, since the isolated measurement points in the coronal slices line up along the z-direction. 
{\color{black}We also presented results for a readout-line-constrained version of LOUPE, which is consistent with widely used 2D acquisition protocols.
These results, we believe, demonstrate the utility and flexibility of LOUPE.}
Nonetheless, integrating the actual physical cost of a specific under-sampling pattern in an MR pulse sequence is a direction that we leave for future research.
One way to achieve this would be to constrain the sub-sampling mask to a class of viable trajectories, for example, by directly parameterizing those trajectories.
{\color{black}Alternatively, one can incorporate sampling constraints that obey a set of predefined hardware requirements (as defined in terms of, e.g., peak currents and maximum slew rates of magnetic gradients), as described recently in~\cite{weiss2019pilot}.}

{\color{black}Our paper restricted its treatment to Cartesian sampling and largely ignored non-Cartesian (off-the-grid) acquisition schemes, which can offer important benefits~\cite{zhang2010real,feng2014golden,lazarus2019sparkling}.
Extension of LOUPE to non-Cartesian settings will involve the use of the non-uniform (inverse) Fourier transform operator or an approximation of it.
Furthermore, one would need to interpolate off-grid measurements in retrospective sampling.
}

{\color{black}Another weakness of the presented LOUPE implementation is due to the relaxation of the thresholding operation, which allowed us to use
a back-propagation-based learning strategy.
However, due to this relaxation, the reconstruction model sees input data that have been rescaled with a continuous value between $0$ and $1$. 
By adopting a relatively large slope parameter ($s=200$), we ensure that these continuous values are almost always very close to zero or 1. 
However, this relaxation means that once we have obtained the ÒoptimizedÓ mask, we need to retrain a reconstruction model with a binary mask. Furthermore, we are likely paying a performance penalty due to the relaxation, since it will undoubtedly influence the optimization. 
We see two ways to address this issue. 
The first approach is to implement the so-called ``straight-through'' gradient estimation strategy, where the continuous relaxation is merely used in the backward pass for computing the gradient and not in the forward pass~\cite{bengio2013estimating}. 
Another approach would involve using non back-propagation based techniques, such as the REINFORCE estimator~\cite{williams1992simple}. 
This technique is known to suffer from high variance and there are several recent papers that have attempted to address this issue~\cite{tucker2017rebar,papini2018stochastic}. 
We plan to explore this direction in future work.}

{\color{black}In this paper, we used an L2 difference between the magnitude of the ground truth and reconstruction images to train LOUPE.
Our experiments with an L2-loss computed on two-channel (complex-valued) reconstructions produced inferior results.
We also noticed that our LOUPE-optimized masks did not emphasize high frequency content, which can also explain some of the blurriness in reconstructions. 
This is particularly noticeable in the line-constrained LOUPE results. 
We believe that our choice of the L2 loss is the main cause of this effect, which can be remedied with alternate loss functions that care more about high-frequency content. }
We note that we experimented with L1 reconstruction loss, as we reported in our conference paper~\cite{bahadir2019learning}.
{\color{black} Our experiments (results not shown) revealed that L1 loss can yield better quality results, particularly as measured by the high frequency error norm (HFEN).}
We intend to explore such differences in future work.
While these loss functions (L1 or L2) are a widely used global metrics of reconstruction quality, they might miss clinically important anatomical details.
An alternative approach could be to devise loss functions that emphasize structural features that are relevant to the clinical application.
This is another future direction we will explore.

In the current version of LOUPE, we did not consider parallel imaging via multiple coils.
Such hardware-based techniques, in combination with compressed sensing, promise to yield even higher degrees of acceleration.
Investigating how to combine LOUPE with multi-coil imaging will be an important area of research, as was recently demonstrated~\cite{gozcu2019rethinking}.

Finally, we would like to emphasize that LOUPE, as in any other data-driven optimization strategy, will perform sub-optimally if test time data differ significantly from training data.
For example, if we train only on healthy subjects and then, at test time, are presented with pathological cases, the reconstructed scans might not be clinically useful.
Thus, carefully monitoring reconstruction quality and ensuring the model is adapted to the new data will be of utmost importance in deploying this approach in the real world.
For example, this could be achieved via fine-tuning the pre-trained LOUPE model on new data, as recently described in~\cite{zhang2019fidelity}.
{\color{black}We also consider machine learning paradigms that don't rely on high quality full-resolution data as an important direction. 
There have been some recent pre-prints in this domain~\cite{liu2019rare,lei2019wasserstein,yaman2019self}.
Integrating such techniques into LOUPE will be critical for certain applications.
For instance, in some anatomical regions such as the abdomen, a fully sampled scenario may not be possible due to organ motion, breathing, the cardiac cycle or other factors. 
This is also true for populations such as young children or patient groups, where subject motion can make the acquisition of high quality fully sampled data impossible. 
}


%

\section*{Acknowledgment}

This work was, in part, supported by NIH R01 grants (R01LM012719 and R01AG053949, to MRS), the NSF NeuroNex grant (1707312, to MRS), and an NSF CAREER grant (1748377, to MRS). This work was also in part supported by a Fulbright Scholarship (to CDB).

Data used in the preparation of this article were obtained from the NYU fastMRI Initiative database~\cite{zbontar2018fastmri}. As such, NYU fastMRI investigators provided data but did not participate in analysis or writing of this report. A listing of NYU fastMRI investigators, subject to updates, can be found at:\url{fastmri.med.nyu.edu}. The primary goal of fastMRI is to test whether machine learning can aid in the reconstruction of medical images.

The concepts and information presented in this paper are based on research results that are not commercially available.

Finally, we would like to express our gratitude for the extremely thorough, thoughtful, and constructive feedback we received from the Associate Editor Justin Haldar, and three anonymous reviewers. 

%
%
%
%
%
%



\bibliographystyle{IEEEtran}
\bibliography{LOUPE-TCI}

\begin{thebibliography}{10}
\providecommand{\url}[1]{#1}
\csname url@samestyle\endcsname
\providecommand{\newblock}{\relax}
\providecommand{\bibinfo}[2]{#2}
\providecommand{\BIBentrySTDinterwordspacing}{\spaceskip=0pt\relax}
\providecommand{\BIBentryALTinterwordstretchfactor}{4}
\providecommand{\BIBentryALTinterwordspacing}{\spaceskip=\fontdimen2\font plus
\BIBentryALTinterwordstretchfactor\fontdimen3\font minus
  \fontdimen4\font\relax}
\providecommand{\BIBforeignlanguage}[2]{{%
\expandafter\ifx\csname l@#1\endcsname\relax
\typeout{** WARNING: IEEEtran.bst: No hyphenation pattern has been}%
\typeout{** loaded for the language `#1'. Using the pattern for}%
\typeout{** the default language instead.}%
\else
\language=\csname l@#1\endcsname
\fi
#2}}
\providecommand{\BIBdecl}{\relax}
\BIBdecl

\bibitem{lustig2008compressed}
M.~Lustig, D.~L. Donoho, J.~M. Santos, and J.~M. Pauly, ``Compressed sensing
  {MRI},'' \emph{{IEEE signal processing magazine}}, vol.~25, no.~2, p.~72,
  2008.

\bibitem{gamper2008compressed}
U.~Gamper, P.~Boesiger, and S.~Kozerke, ``Compressed sensing in dynamic
  {MRI},'' \emph{{Magnetic Resonance in Medicine}}, vol.~59, no.~2, pp.
  365--373, 2008.

\bibitem{ma2008efficient}
S.~Ma, W.~Yin, Y.~Zhang, and A.~Chakraborty, ``An efficient algorithm for
  compressed {MR} imaging using total variation and wavelets,'' in \emph{2008
  {IEEE Conference on Computer Vision and Pattern Recognition}}.\hskip 1em plus
  0.5em minus 0.4em\relax IEEE, 2008, pp. 1--8.

\bibitem{qu2014magnetic}
X.~Qu, Y.~Hou, F.~Lam, D.~Guo, J.~Zhong, and Z.~Chen, ``Magnetic resonance
  image reconstruction from undersampled measurements using a patch-based
  nonlocal operator,'' \emph{Medical image analysis}, vol.~18, no.~6, pp.
  843--856, 2014.

\bibitem{ravishankar2011mr}
S.~Ravishankar and Y.~Bresler, ``{MR} image reconstruction from highly
  undersampled k-space data by dictionary learning,'' \emph{{IEEE Transactions
  on Medical Imaging}}, vol.~30, no.~5, pp. 1028--1041, 2011.

\bibitem{zhan2016fast}
Z.~Zhan, J.-F. Cai, D.~Guo, Y.~Liu, Z.~Chen, and X.~Qu, ``Fast multiclass
  dictionaries learning with geometrical directions in {MRI} reconstruction,''
  \emph{{IEEE Transactions on Biomedical Engineering}}, vol.~63, no.~9, pp.
  1850--1861, 2016.

\bibitem{wang2010variable}
Z.~Wang and G.~R. Arce, ``Variable density compressed image sampling,''
  \emph{{IEEE Transactions on Image Processing}}, vol.~19, no.~1, pp. 264--270,
  2010.

\bibitem{haldar2011compressed}
J.~P. Haldar, D.~Hernando, and Z.-P. Liang, ``Compressed-sensing {MRI} with
  random encoding,'' \emph{{IEEE Transactions on Medical Imaging}}, vol.~30,
  no.~4, pp. 893--903, 2011.

\bibitem{bahadir2019learning}
C.~D. Bahadir, A.~V. Dalca, and M.~R. Sabuncu, ``Learning-based optimization of
  the under-sampling pattern in {MRI},'' \emph{Proc of {Information Processing
  in Medical Imaging}}, 2019.

\bibitem{sun2016deep}
J.~Sun, H.~Li, Z.~Xu \emph{et~al.}, ``Deep {ADMM-Net for compressive sensing
  MRI},'' in \emph{Advances in neural information processing systems}, 2016,
  pp. 10--18.

\bibitem{tezcan2018mr}
K.~C. Tezcan, C.~F. Baumgartner, R.~Luechinger, K.~P. Pruessmann, and
  E.~Konukoglu, ``{MR} image reconstruction using deep density priors,''
  \emph{IEEE transactions on medical imaging}, 2018.

\bibitem{wang2018image}
G.~Wang, J.~C. Ye, K.~Mueller, and J.~A. Fessler, ``Image reconstruction is a
  new frontier of machine learning,'' \emph{IEEE transactions on medical
  imaging}, vol.~37, no.~6, pp. 1289--1296, 2018.

\bibitem{zhu2018image}
B.~Zhu, J.~Z. Liu, S.~F. Cauley, B.~R. Rosen, and M.~S. Rosen, ``Image
  reconstruction by domain-transform manifold learning,'' \emph{Nature}, vol.
  555, no. 7697, p. 487, 2018.

\bibitem{aggarwal2018modl}
H.~K. Aggarwal, M.~P. Mani, and M.~Jacob, ``Modl: Model-based deep learning
  architecture for inverse problems,'' \emph{IEEE transactions on medical
  imaging}, vol.~38, no.~2, pp. 394--405, 2018.

\bibitem{hammernik2018learning}
K.~Hammernik, T.~Klatzer, E.~Kobler, M.~P. Recht, D.~K. Sodickson, T.~Pock, and
  F.~Knoll, ``Learning a variational network for reconstruction of accelerated
  {MRI} data,'' \emph{Magnetic resonance in medicine}, vol.~79, no.~6, pp.
  3055--3071, 2018.

\bibitem{huang2014bayesian}
Y.~Huang, J.~Paisley, Q.~Lin, X.~Ding, X.~Fu, and X.-P. Zhang, ``Bayesian
  nonparametric dictionary learning for compressed sensing {MRI},'' \emph{{IEEE
  Transactions on Image Processing}}, vol.~23, no.~12, pp. 5007--5019, 2014.

\bibitem{yang2017admm}
Y.~Yang, J.~Sun, H.~Li, and Z.~Xu, ``{ADMM-Net: A deep learning approach for
  compressive sensing MRI},'' \emph{arXiv preprint arXiv:1705.06869}, 2017.

\bibitem{ronneberger2015u}
O.~Ronneberger, P.~Fischer, and T.~Brox, ``U-net: Convolutional networks for
  biomedical image segmentation,'' in \emph{International Conference on Medical
  image computing and computer-assisted intervention}.\hskip 1em plus 0.5em
  minus 0.4em\relax Springer, 2015, pp. 234--241.

\bibitem{lee2017deep}
D.~Lee, J.~Yoo, and J.~C. Ye, ``Deep residual learning for compressed sensing
  {MRI},'' in \emph{2017 {IEEE 14th International Symposium on Biomedical
  Imaging (ISBI 2017)}}.\hskip 1em plus 0.5em minus 0.4em\relax IEEE, 2017, pp.
  15--18.

\bibitem{hyun2018deep}
C.~M. Hyun, H.~P. Kim, S.~M. Lee, S.~Lee, and J.~K. Seo, ``Deep learning for
  undersampled {MRI} reconstruction,'' \emph{Physics in Medicine \& Biology},
  vol.~63, no.~13, p. 135007, 2018.

\bibitem{wu2019phasecam3d}
Y.~Wu, V.~Boominathan, H.~Chen, A.~Sankaranarayanan, and A.~Veeraraghavan,
  ``Phasecam3d—learning phase masks for passive single view depth
  estimation,'' in \emph{2019 IEEE International Conference on Computational
  Photography (ICCP)}.\hskip 1em plus 0.5em minus 0.4em\relax IEEE, 2019, pp.
  1--12.

\bibitem{goodfellow2014generative}
I.~Goodfellow, J.~Pouget-Abadie, M.~Mirza, B.~Xu, D.~Warde-Farley, S.~Ozair,
  A.~Courville, and Y.~Bengio, ``Generative adversarial nets,'' in
  \emph{Advances in neural information processing systems}, 2014, pp.
  2672--2680.

\bibitem{yang2018dagan}
G.~Yang, S.~Yu, H.~Dong, G.~Slabaugh, P.~L. Dragotti, X.~Ye, F.~Liu,
  S.~Arridge, J.~Keegan, Y.~Guo \emph{et~al.}, ``{DAGAN: deep de-aliasing
  generative adversarial networks for fast compressed sensing MRI
  reconstruction},'' \emph{{IEEE Transactions on Medical Imaging}}, vol.~37,
  no.~6, pp. 1310--1321, 2018.

\bibitem{quan2018compressed}
T.~M. Quan, T.~Nguyen-Duc, and W.-K. Jeong, ``Compressed sensing mri
  reconstruction using a generative adversarial network with a cyclic loss,''
  \emph{{IEEE Transactions on Medical Imaging}}, vol.~37, no.~6, pp.
  1488--1497, 2018.

\bibitem{mardani2017deep}
M.~Mardani, E.~Gong, J.~Y. Cheng, S.~Vasanawala, G.~Zaharchuk, M.~Alley,
  N.~Thakur, S.~Han, W.~Dally, J.~M. Pauly \emph{et~al.}, ``Deep generative
  adversarial networks for compressed sensing automates {MRI},'' \emph{arXiv
  preprint arXiv:1706.00051}, 2017.

\bibitem{lei2019wasserstein}
K.~Lei, M.~Mardani, J.~M. Pauly, and S.~S. Vasawanala, ``Wasserstein {GAN}s for
  {MR} imaging: from paired to unpaired training,'' \emph{arXiv preprint
  arXiv:1910.07048}, 2019.

\bibitem{lustig2007sparse}
M.~Lustig, D.~Donoho, and J.~M. Pauly, ``Sparse {MRI:} the application of
  compressed sensing for rapid mr imaging,'' \emph{Magnetic Resonance in
  Medicine: An Official Journal of the International Society for Magnetic
  Resonance in Medicine}, vol.~58, no.~6, pp. 1182--1195, 2007.

\bibitem{elad2007optimized}
M.~Elad, ``Optimized projections for compressed sensing,'' \emph{{IEEE
  Transactions on Signal Processing}}, vol.~55, no.~12, pp. 5695--5702, 2007.

\bibitem{xu2010optimized}
J.~Xu, Y.~Pi, and Z.~Cao, ``Optimized projection matrix for compressive
  sensing,'' \emph{EURASIP Journal on Advances in Signal Processing}, vol.
  2010, no.~1, p. 560349, 2010.

\bibitem{puy2011variable}
G.~Puy, P.~Vandergheynst, and Y.~Wiaux, ``On variable density compressive
  sampling,'' \emph{{IEEE Signal Processing Letters}}, vol.~18, no.~10, pp.
  595--598, 2011.

\bibitem{li2013projection}
G.~Li, Z.~Zhu, D.~Yang, L.~Chang, and H.~Bai, ``On projection matrix
  optimization for compressive sensing systems,'' \emph{IEEE Transactions on
  Signal Processing}, vol.~61, no.~11, pp. 2887--2898, 2013.

\bibitem{knoll2011adapted}
F.~Knoll, C.~Clason, C.~Diwoky, and R.~Stollberger, ``Adapted random sampling
  patterns for accelerated {MRI},'' \emph{Magnetic resonance materials in
  physics, biology and medicine}, vol.~24, no.~1, pp. 43--50, 2011.

\bibitem{kumar2008durga}
C.~Kumar~Anand, A.~Thomas~Curtis, and R.~Kumar, ``Durga: A
  heuristically-optimized data collection strategy for volumetric magnetic
  resonance imaging,'' \emph{Engineering Optimization}, vol.~40, no.~2, pp.
  117--136, 2008.

\bibitem{curtis2008random}
A.~T. Curtis and C.~K. Anand, ``Random volumetric mri trajectories via genetic
  algorithms,'' \emph{Journal of Biomedical Imaging}, vol. 2008, p.~6, 2008.

\bibitem{seeger2010optimization}
M.~Seeger, H.~Nickisch, R.~Pohmann, and B.~Sch{\"o}lkopf, ``Optimization of
  k-space trajectories for compressed sensing by {Bayesian} experimental
  design,'' \emph{{Magnetic Resonance in Medicine}}, vol.~63, no.~1, pp.
  116--126, 2010.

\bibitem{haldar2019oedipus}
J.~P. Haldar and D.~Kim, ``{OEDIPUS:} an experiment design framework for
  sparsity-constrained {MRI},'' \emph{{IEEE Transactions on Medical Imaging}},
  2019.

\bibitem{roman2014asymptotic}
B.~Roman, A.~Hansen, and B.~Adcock, ``On asymptotic structure in compressed
  sensing,'' \emph{arXiv preprint arXiv:1406.4178}, 2014.

\bibitem{sherry2019learning}
F.~Sherry, M.~Benning, J.~C. D.~l. Reyes, M.~J. Graves, G.~Maierhofer,
  G.~Williams, C.-B. Sch{\"o}nlieb, and M.~J. Ehrhardt, ``Learning the sampling
  pattern for mri,'' \emph{arXiv preprint arXiv:1906.08754}, 2019.

\bibitem{ravishankar2011adaptive}
S.~Ravishankar and Y.~Bresler, ``Adaptive sampling design for compressed
  sensing {MRI},'' in \emph{2011 Annual International Conference of the IEEE
  Engineering in Medicine and Biology Society}.\hskip 1em plus 0.5em minus
  0.4em\relax IEEE, 2011, pp. 3751--3755.

\bibitem{liu2012under}
D.-d. Liu, D.~Liang, X.~Liu, and Y.-t. Zhang, ``Under-sampling trajectory
  design for compressed sensing {MRI},'' in \emph{2012 Annual International
  Conference of the IEEE Engineering in Medicine and Biology Society}.\hskip
  1em plus 0.5em minus 0.4em\relax IEEE, 2012, pp. 73--76.

\bibitem{baldassarre2016learning}
L.~Baldassarre, Y.-H. Li, J.~Scarlett, B.~G{\"o}zc{\"u}, I.~Bogunovic, and
  V.~Cevher, ``Learning-based compressive subsampling,'' \emph{{IEEE} Journal
  of Selected Topics in Signal Processing}, vol.~10, no.~4, pp. 809--822, 2016.

\bibitem{gozcu2018learning}
B.~G{\"o}zc{\"u}, R.~K. Mahabadi, Y.-H. Li, E.~Il{\i}cak, T.~Cukur,
  J.~Scarlett, and V.~Cevher, ``Learning-based compressive {MRI},'' \emph{{IEEE
  Transactions on Medical Imaging}}, vol.~37, no.~6, pp. 1394--1406, 2018.

\bibitem{sanchez2020scalable}
T.~Sanchez, B.~G{\"o}zc{\"u}, R.~B. van Heeswijk, A.~Eftekhari, E.~Il{\i}cak,
  T.~{\c{C}}ukur, and V.~Cevher, ``Scalable learning-based sampling
  optimization for compressive dynamic {MRI},'' in \emph{ICASSP 2020-2020 IEEE
  International Conference on Acoustics, Speech and Signal Processing
  (ICASSP)}.\hskip 1em plus 0.5em minus 0.4em\relax IEEE, 2020, pp. 8584--8588.

\bibitem{zijlstra2016evaluation}
F.~Zijlstra, M.~A. Viergever, and P.~R. Seevinck, ``Evaluation of variable
  density and data-driven k-space undersampling for compressed sensing magnetic
  resonance imaging,'' \emph{Investigative radiology}, vol.~51, no.~6, pp.
  410--419, 2016.

\bibitem{weiss2019learning}
T.~Weiss, S.~Vedula, O.~Senouf, A.~Bronstein, O.~Michailovich, and
  M.~Zibulevsky, ``Learning fast magnetic resonance imaging,'' \emph{arXiv
  preprint arXiv:1905.09324}, 2019.

\bibitem{aggarwal2019joint}
H.~K. Aggarwal and M.~Jacob, ``Joint optimization of sampling patterns and deep
  priors for improved parallel mri,'' \emph{arXiv preprint arXiv:1911.02945},
  2019.

\bibitem{huijben2020learning}
I.~A. Huijben, B.~S. Veeling, and R.~J. van Sloun, ``Learning sampling and
  model-based signal recovery for compressed sensing {MRI},'' in \emph{ICASSP
  2020-2020 IEEE International Conference on Acoustics, Speech and Signal
  Processing (ICASSP)}.\hskip 1em plus 0.5em minus 0.4em\relax IEEE, 2020, pp.
  8906--8910.

\bibitem{kingma2013auto}
D.~P. Kingma and M.~Welling, ``Auto-encoding variational {Bayes},'' \emph{arXiv
  preprint arXiv:1312.6114}, 2013.

\bibitem{jang2016categorical}
E.~Jang, S.~Gu, and B.~Poole, ``Categorical reparameterization with
  {Gumbel-softmax},'' \emph{arXiv preprint arXiv:1611.01144}, 2016.

\bibitem{maddison2016concrete}
C.~J. Maddison, A.~Mnih, and Y.~W. Teh, ``The concrete distribution: A
  continuous relaxation of discrete random variables,'' \emph{arXiv preprint
  arXiv:1611.00712}, 2016.

\bibitem{schlemper2017deep}
J.~Schlemper, J.~Caballero, J.~V. Hajnal, A.~Price, and D.~Rueckert, ``A deep
  cascade of convolutional neural networks for {MR} image reconstruction,'' in
  \emph{International Conference on Information Processing in Medical
  Imaging}.\hskip 1em plus 0.5em minus 0.4em\relax Springer, 2017, pp.
  647--658.

\bibitem{chollet2015keras}
F.~Chollet \emph{et~al.}, ``Keras,'' 2015.

\bibitem{abadi2016tensorflow}
M.~Abadi, P.~Barham, J.~Chen, Z.~Chen, A.~Davis, J.~Dean, M.~Devin,
  S.~Ghemawat, G.~Irving, M.~Isard \emph{et~al.}, ``Tensorflow: A system for
  large-scale machine learning,'' in \emph{12th $\{$USENIX$\}$ Symposium on
  Operating Systems Design and Implementation ($\{$OSDI$\}$ 16)}, 2016, pp.
  265--283.

\bibitem{dalca2018anatomical}
A.~V. Dalca, J.~Guttag, and M.~R. Sabuncu, ``Anatomical priors in convolutional
  networks for unsupervised biomedical segmentation,'' in \emph{{Proceedings of
  the IEEE Conference on Computer Vision and Pattern Recognition}}, 2018, pp.
  9290--9299.

\bibitem{kingma2014adam}
D.~P. Kingma and J.~Ba, ``Adam: A method for stochastic optimization,''
  \emph{arXiv preprint arXiv:1412.6980}, 2014.

\bibitem{zbontar2018fastmri}
J.~Zbontar, F.~Knoll, A.~Sriram, M.~J. Muckley, M.~Bruno, A.~Defazio,
  M.~Parente, K.~J. Geras, J.~Katsnelson, H.~Chandarana, Z.~Zhang, M.~Drozdzal,
  A.~Romero, M.~Rabbat, P.~Vincent, J.~Pinkerton, D.~Wang, N.~Yakubova,
  E.~Owens, C.~L. Zitnick, M.~P. Recht, D.~K. Sodickson, and Y.~W. Lui,
  ``{FastMRI: An open dataset and benchmarks for accelerated MRI},''
  \emph{arXiv preprint arXiv:1811.08839}, 2018.

\bibitem{welstead1999fractal}
S.~T. Welstead, \emph{Fractal and wavelet image compression techniques}.\hskip
  1em plus 0.5em minus 0.4em\relax {SPIE} Optical Engineering Press Bellingham,
  Washington, 1999.

\bibitem{wang2004image}
Z.~Wang, A.~C. Bovik, H.~R. Sheikh, E.~P. Simoncelli \emph{et~al.}, ``Image
  quality assessment: from error visibility to structural similarity,''
  \emph{{IEEE Transactions on Image Processing}}, vol.~13, no.~4, pp. 600--612,
  2004.

\bibitem{eksioglu2016decoupled}
E.~M. Eksioglu, ``Decoupled algorithm for {MRI} reconstruction using nonlocal
  block matching model: {BM3D-MRI},'' \emph{Journal of Mathematical Imaging and
  Vision}, vol.~56, no.~3, pp. 430--440, 2016.

\bibitem{haldar2014low}
J.~P. Haldar, ``{Low-rank modeling of local $ k $-space neighborhoods (LORAKS)
  for constrained MRI},'' \emph{{IEEE Transactions on Medical Imaging}},
  vol.~33, no.~3, pp. 668--681, 2014.

\bibitem{haldar2016p}
J.~P. Haldar and J.~Zhuo, ``{P-LORAKS: Low-rank modeling of local k-space
  neighborhoods with parallel imaging data},'' \emph{{Magnetic Resonance in
  Medicine}}, vol.~75, no.~4, pp. 1499--1514, 2016.

\bibitem{guo2014new}
W.~Guo, J.~Qin, and W.~Yin, ``A new detail-preserving regularization scheme,''
  \emph{{SIAM Journal on Imaging Sciences}}, vol.~7, no.~2, pp. 1309--1334,
  2014.

\bibitem{vellagoundar2015robust}
J.~Vellagoundar and R.~R. Machireddy, ``A robust adaptive sampling method for
  faster acquisition of {MR} images,'' \emph{Magnetic resonance imaging},
  vol.~33, no.~5, pp. 635--643, 2015.

\bibitem{di2014autism}
A.~Di~Martino, C.-G. Yan, Q.~Li, E.~Denio, F.~X. Castellanos, K.~Alaerts, J.~S.
  Anderson, M.~Assaf, S.~Y. Bookheimer, M.~Dapretto \emph{et~al.}, ``The autism
  brain imaging data exchange: towards a large-scale evaluation of the
  intrinsic brain architecture in autism,'' \emph{Molecular psychiatry},
  vol.~19, no.~6, p. 659, 2014.

\bibitem{ramzi2020benchmarking}
Z.~Ramzi, P.~Ciuciu, and J.-L. Starck, ``Benchmarking {MRI} reconstruction
  neural networks on large public datasets,'' \emph{Applied Sciences}, vol.~10,
  no.~5, p. 1816, 2020.

\bibitem{weiss2019pilot}
T.~Weiss, O.~Senouf, S.~Vedula, O.~Michailovich, M.~Zibulevsky, and
  A.~Bronstein, ``Pilot: Physics-informed learned optimal trajectories for
  accelerated {MRI},'' \emph{arXiv preprint arXiv:1909.05773}, 2019.

\bibitem{zhang2010real}
S.~Zhang, M.~Uecker, D.~Voit, K.-D. Merboldt, and J.~Frahm, ``Real-time
  cardiovascular magnetic resonance at high temporal resolution: radial {FLASH}
  with nonlinear inverse reconstruction,'' \emph{Journal of Cardiovascular
  Magnetic Resonance}, vol.~12, no.~1, p.~39, 2010.

\bibitem{feng2014golden}
L.~Feng, R.~Grimm, K.~T. Block, H.~Chandarana, S.~Kim, J.~Xu, L.~Axel, D.~K.
  Sodickson, and R.~Otazo, ``Golden-angle radial sparse parallel {MRI:}
  combination of compressed sensing, parallel imaging, and golden-angle radial
  sampling for fast and flexible dynamic volumetric {MRI},'' \emph{Magnetic
  resonance in medicine}, vol.~72, no.~3, pp. 707--717, 2014.

\bibitem{lazarus2019sparkling}
C.~Lazarus, P.~Weiss, N.~Chauffert, F.~Mauconduit, L.~El~Gueddari,
  C.~Destrieux, I.~Zemmoura, A.~Vignaud, and P.~Ciuciu, ``{SPARKLING:
  variable-density k-space filling curves for accelerated T2*-weighted MRI},''
  \emph{Magnetic resonance in medicine}, vol.~81, no.~6, pp. 3643--3661, 2019.

\bibitem{bengio2013estimating}
Y.~Bengio, N.~L{\'e}onard, and A.~Courville, ``Estimating or propagating
  gradients through stochastic neurons for conditional computation,''
  \emph{arXiv preprint arXiv:1308.3432}, 2013.

\bibitem{williams1992simple}
R.~J. Williams, ``Simple statistical gradient-following algorithms for
  connectionist reinforcement learning,'' \emph{Machine learning}, vol.~8, no.
  3-4, pp. 229--256, 1992.

\bibitem{tucker2017rebar}
G.~Tucker, A.~Mnih, C.~J. Maddison, J.~Lawson, and J.~Sohl-Dickstein, ``Rebar:
  Low-variance, unbiased gradient estimates for discrete latent variable
  models,'' in \emph{Advances in Neural Information Processing Systems}, 2017,
  pp. 2627--2636.

\bibitem{papini2018stochastic}
M.~Papini, D.~Binaghi, G.~Canonaco, M.~Pirotta, and M.~Restelli, ``Stochastic
  variance-reduced policy gradient,'' \emph{arXiv preprint arXiv:1806.05618},
  2018.

\bibitem{gozcu2019rethinking}
B.~G{\"o}zc{\"u}, T.~Sanchez, and V.~Cevher, ``Rethinking sampling in parallel
  {MRI}: A data-driven approach,'' in \emph{2019 27th European Signal
  Processing Conference (EUSIPCO)}.\hskip 1em plus 0.5em minus 0.4em\relax
  IEEE, 2019, pp. 1--5.

\bibitem{zhang2019fidelity}
J.~Zhang, Z.~Liu, S.~Zhang, H.~Zhang, P.~Spincemaille, T.~D. Nguyen, M.~R.
  Sabuncu, and Y.~Wang, ``Fidelity imposed network edit {(FINE)} for solving
  ill-posed image reconstruction,'' \emph{arXiv preprint arXiv:1905.07284},
  2019.

\bibitem{liu2019rare}
J.~Liu, Y.~Sun, C.~Eldeniz, W.~Gan, H.~An, and U.~S. Kamilov, ``{RARE}: Image
  reconstruction using deep priors learned without ground truth,'' \emph{arXiv
  preprint arXiv:1912.05854}, 2019.

\bibitem{yaman2019self}
B.~Yaman, S.~A.~H. Hosseini, S.~Moeller, J.~Ellermann, K.~U{\u{g}}urbil, and
  M.~Ak{\c{c}}akaya, ``Self-supervised learning of physics-based reconstruction
  neural networks without fully-sampled reference data,'' \emph{arXiv preprint
  arXiv:1912.07669}, 2019.

\end{thebibliography}

\newpage

\section{Supplementary Material}
\setcounter{figure}{0}

\makeatletter 
\renewcommand{\thefigure}{S\@arabic\c@figure}
\makeatother

\subsection{Fine-tuning the slope hyperparameters}

\begin{figure*}[h]
\begin{centering}
\includegraphics[width=10cm]{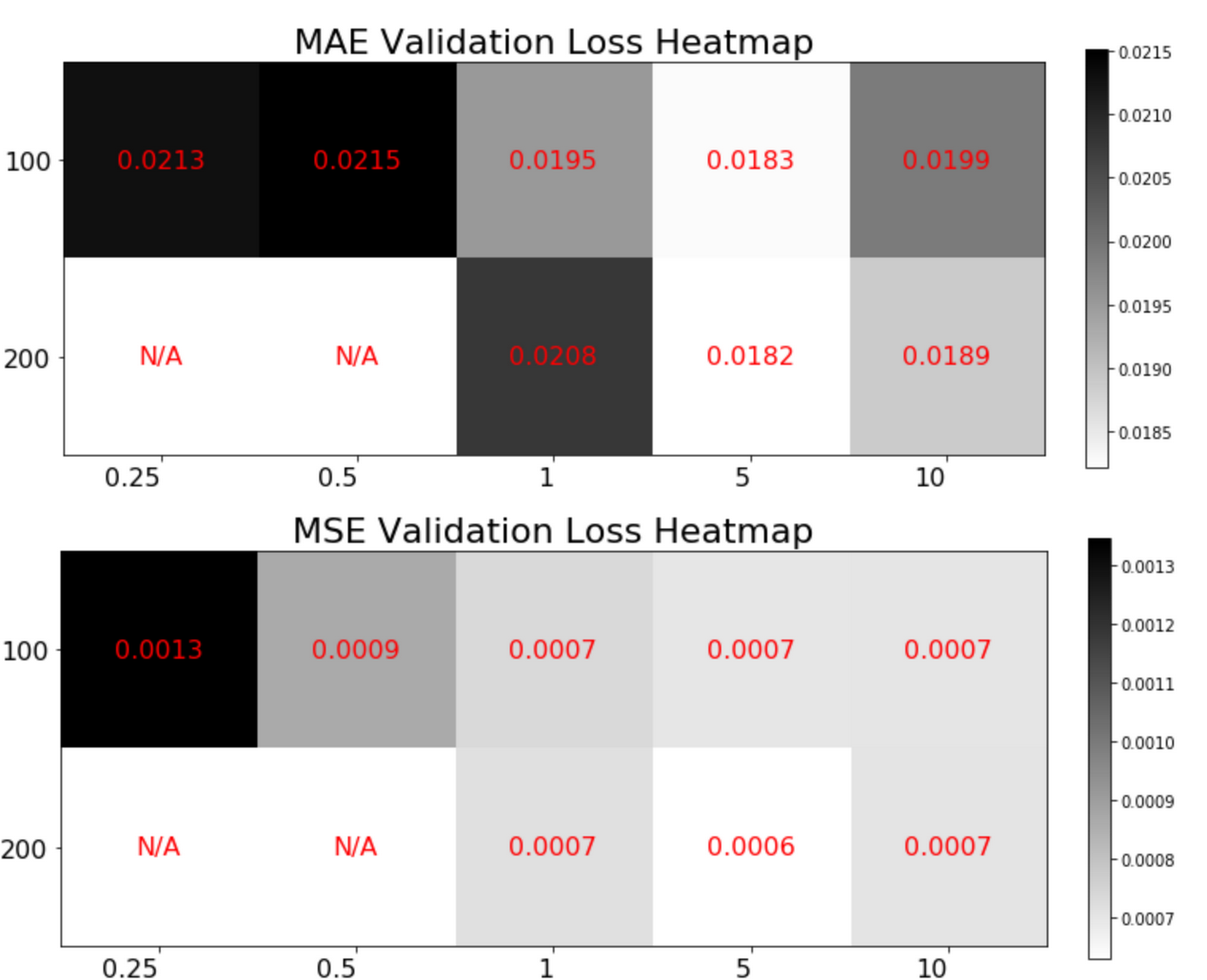}
\caption{Grid search for the slope hyper-parameters $s$ and $t$, where MAE (L1 loss) and MSE (L2 loss) on the validation data are listed. N/A indicates failure of optimization due to numerical issues.} 
\label{fig:slope-tune}
\end{centering}
\end{figure*}

Figure~\ref{fig:slope-tune} shows the validation set MAE and MSE for LOUPE with masks optimized using a variety of hyper-parameter values for the two sigmoidal slopes $s$ and $t$.
These errors are relatively stable compared to the variation of results among the baselines in Table I. 
For example, for MAE, changing $s$ by ten-fold and $t$ by 2-fold results in a loss variation of less than $\pm0.02$, which is well within the variability of the results produced by the benchmark masks. 

\subsection{Details of the Cascade-Net Implementation}
We implemented the identical architecture described in the original Cascade-Net paper. 
Specifically, for the regularization layers of Cascade-Net, a 5-layer model was used with channel size 64 at each layer. Each layer consists of convolution followed by a ReLU activation function. We used a residual learning strategy which adds the zero-filled input to the output of the CNN. The overall architecture is unrolled such that K=5. In the original paper and in our case, lambda is not learnable but is a hyperparameter that we set to $0.001$, which was found by grid search over validation loss and further optimized using a Bayesian Optimization hyper-parameter tuning package: {\url{https://iopscience.iop.org/article/10.1088/1749-4699/8/1/014008/meta}}.

\newpage

\begin{figure*}[ht]
\begin{centering}
\includegraphics[width=10cm]{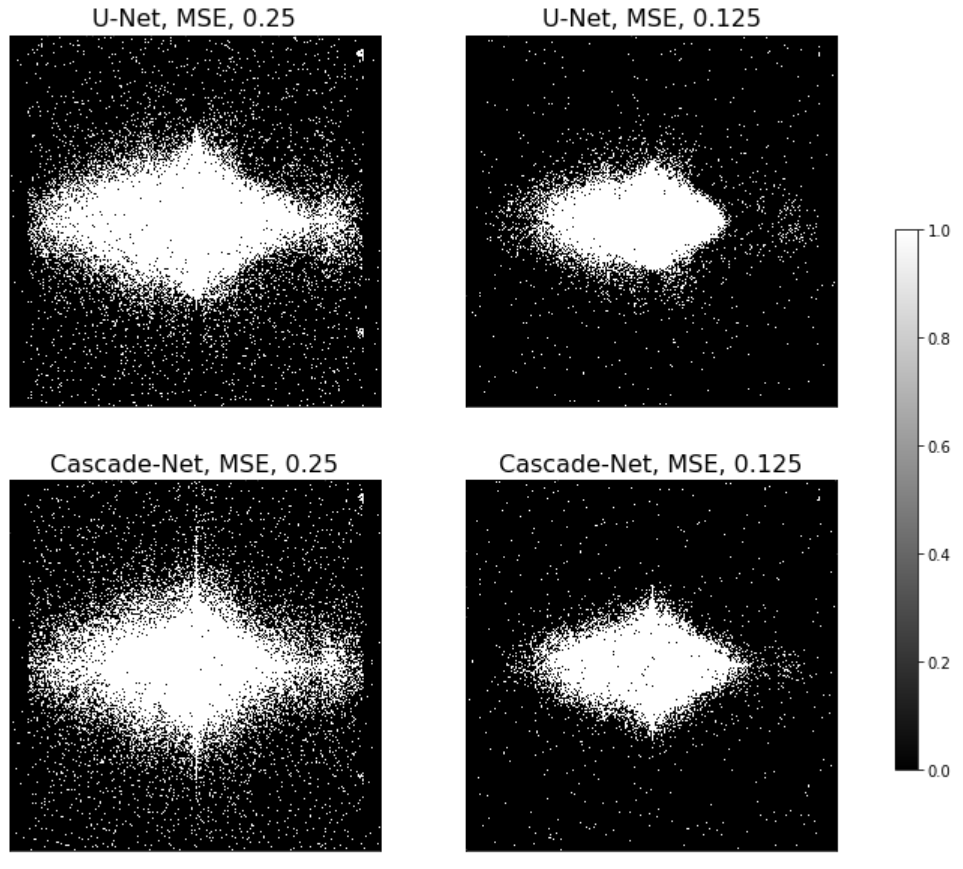}
\caption{Learned LOUPE masks using NYU fastMRI dataset with 4-fold ($\alpha=0.25$) and 8-fold ($\alpha=0.125$)acceleration rates.
Two versions of LOUPE are compared. 
One with U-Net as its reconstruction network (top row), and the other with Cascade-Net (bottom row).} 
\label{fig:mask-compare-4x}
\end{centering}
\end{figure*}

\begin{figure*}[h]
\begin{centering}
\includegraphics[width=10cm]{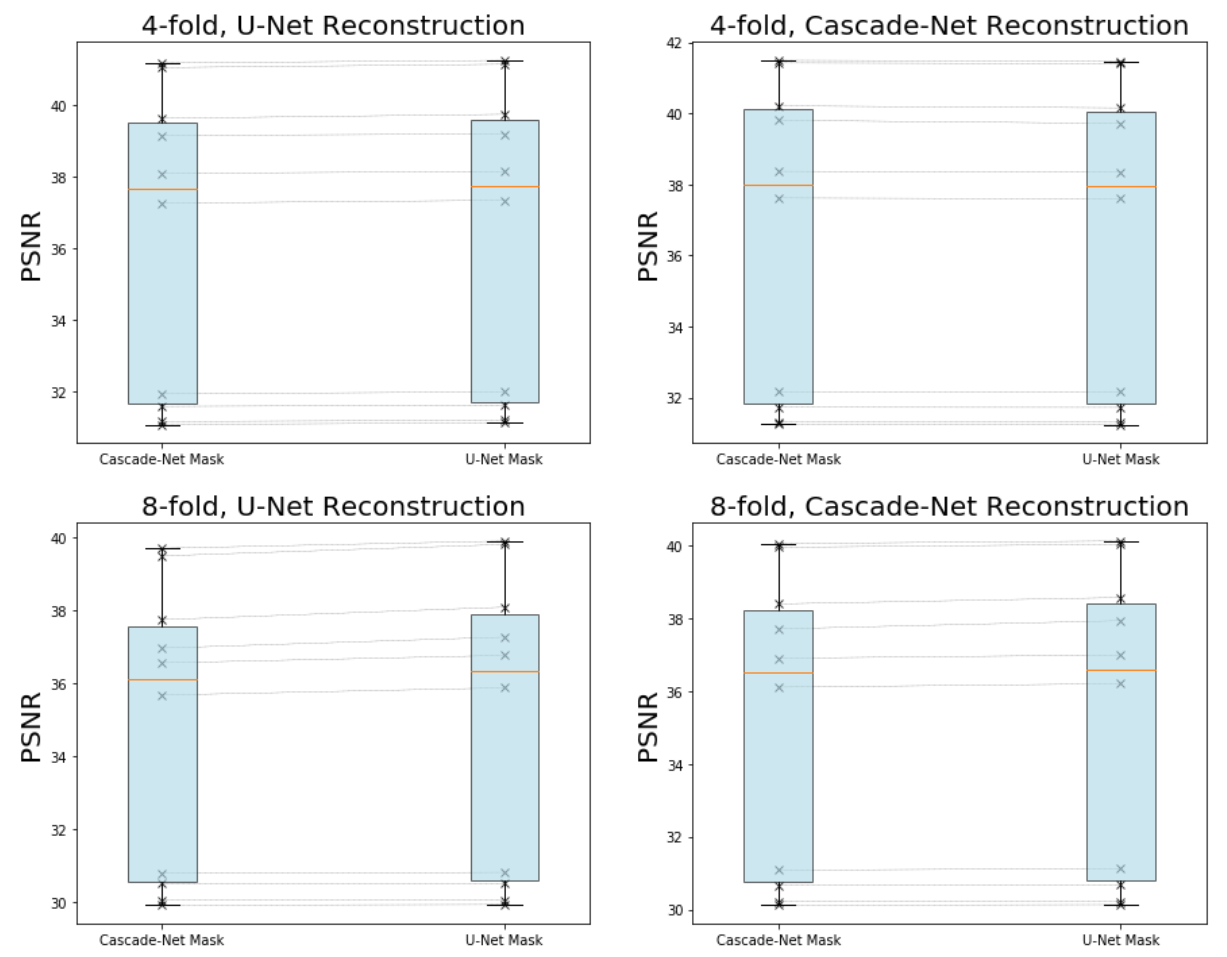}
\includegraphics[width=10cm]{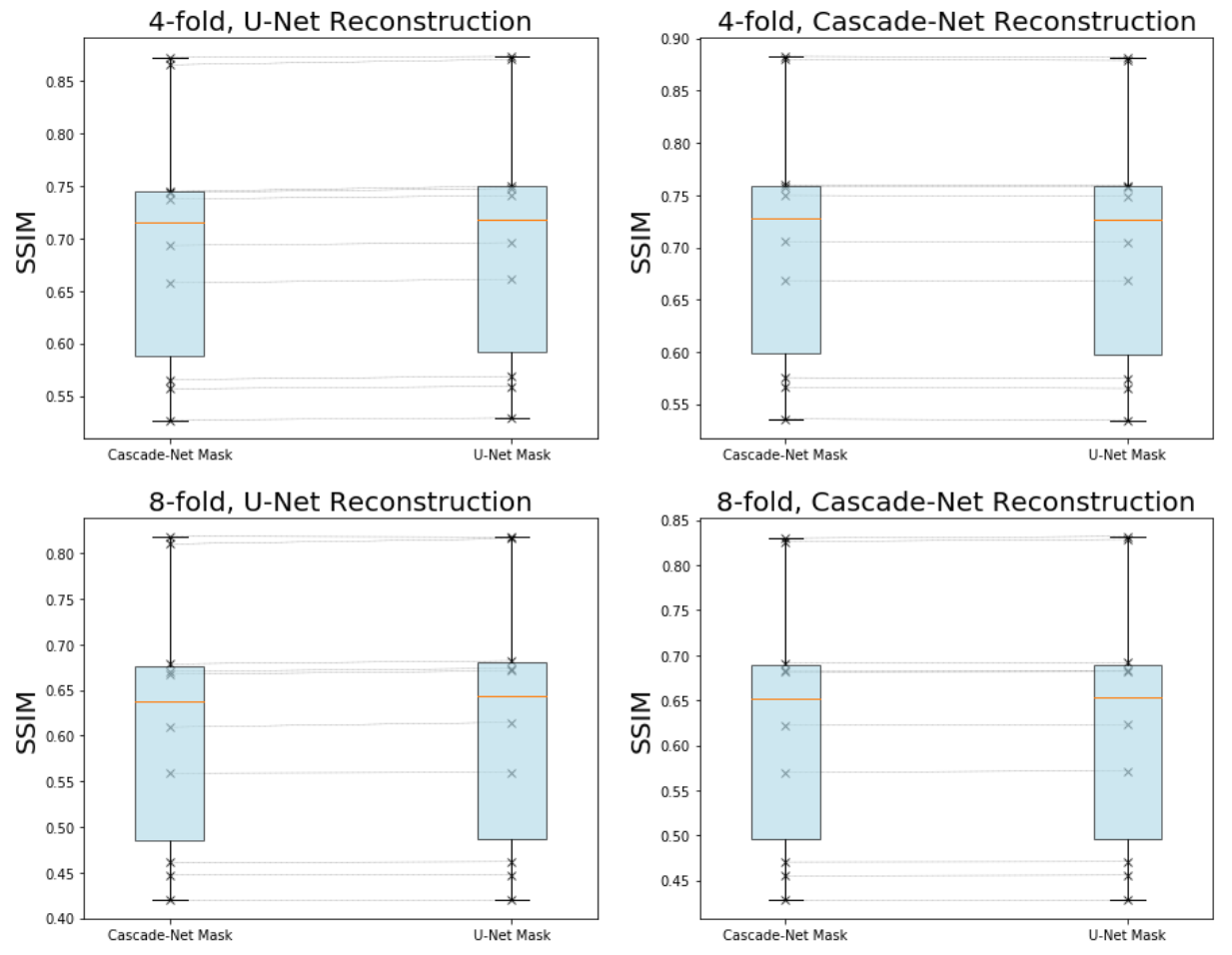}
\includegraphics[width=10cm]{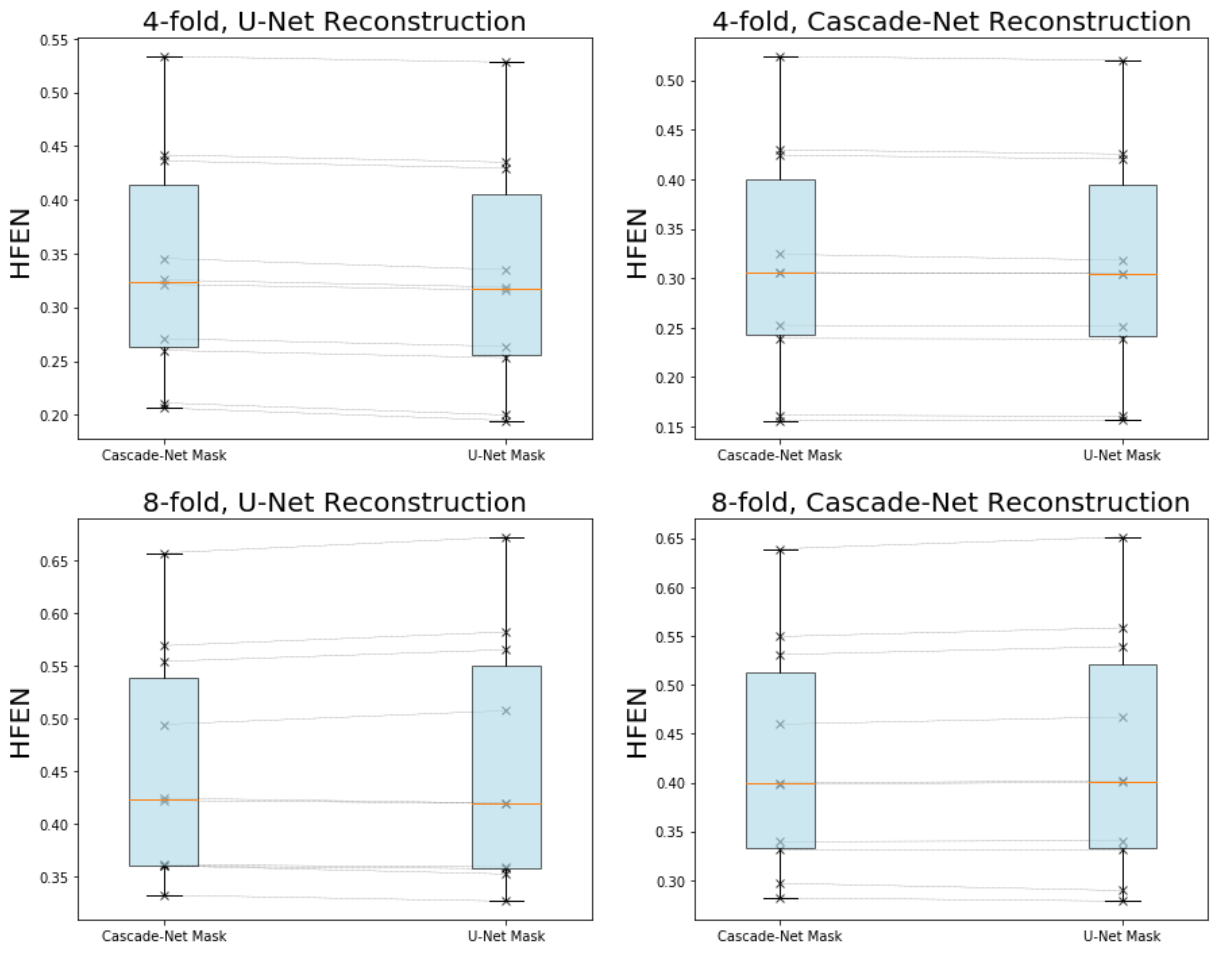}
\caption{Reconstruction performance of U-Net and Cascade-Net models with different LOUPE optimized masks and acceleration rates on the test knee scans. Masks were computed with the two different versions of LOUPE - one with U-Net as its reconstruction network, and the other with Cascade-Net.
We observe that the performance metrics are very close between different masks and reconstruction networks.} 
\label{fig:mask-compare-4x}
\end{centering}
\end{figure*}

\begin{sidewaystable*}[t]
\tiny
\caption{Supplementary Table, Mean Values $\pm$ Standard Deviations} \label{table:table_additional_pd} 
\begin{tabular}{lllllllllllll}
Acceleration             & Method & Masks                  & MSE PD            & MSE PDFS          & MAE PD          & MAE PDFS        & HFEN PD         & HFEN PDFS       & PSNR PD      & PSNR PDFS    & SSIM PD     & SSIM PDFS   \\ \hline
\multirow{24}{*}{4-fold} & BM3D   & Cartesian              & 0.00796 $\pm$ 0.00176 & 0.00629 $\pm$ 0.00127 & 0.0646 $\pm$ 0.0092 & 0.0587 $\pm$ 0.0051 & 0.9484 $\pm$ 0.0091 & 1.0596 $\pm$ 0.0199 & 21.1 $\pm$ 0.99  & 22.1 $\pm$ 0.9   & 0.59 $\pm$ 0.03 & 0.63 $\pm$ 0.01 \\
                         & BM3D   & LOUPE Line Constrained & 0.00023 $\pm$ 8e-05   & 0.00089 $\pm$ 7e-05   & 0.0105 $\pm$ 0.0018 & 0.0234 $\pm$ 0.001  & 0.548 $\pm$ 0.0421  & 0.7096 $\pm$ 0.0715 & 36.69 $\pm$ 1.62 & 30.5 $\pm$ 0.36  & 0.95 $\pm$ 0.02 & 0.82 $\pm$ 0.02 \\
                         & BM3D   & Uniform Random         & 0.01074 $\pm$ 0.0035  & 0.00902 $\pm$ 0.00081 & 0.0707 $\pm$ 0.0176 & 0.0695 $\pm$ 0.0063 & 0.9467 $\pm$ 0.0738 & 1.1416 $\pm$ 0.1027 & 19.95 $\pm$ 1.56 & 20.47 $\pm$ 0.39 & 0.63 $\pm$ 0.06 & 0.58 $\pm$ 0.02 \\
                         & BM3D   & Variable Density       & 0.00032 $\pm$ 0.00011 & 0.00137 $\pm$ 0.00013 & 0.0128 $\pm$ 0.0023 & 0.029 $\pm$ 0.0013  & 0.5772 $\pm$ 0.0449 & 0.8288 $\pm$ 0.0846 & 35.2 $\pm$ 1.54  & 28.66 $\pm$ 0.41 & 0.93 $\pm$ 0.02 & 0.75 $\pm$ 0.02 \\
                         & BM3D   & Spectrum               & 0.00021 $\pm$ 9e-05   & 0.00135 $\pm$ 0.00012 & 0.0107 $\pm$ 0.0021 & 0.0287 $\pm$ 0.0013 & 0.537 $\pm$ 0.0464  & 0.8464 $\pm$ 0.0831 & 37.05 $\pm$ 1.78 & 28.72 $\pm$ 0.4  & 0.94 $\pm$ 0.02 & 0.75 $\pm$ 0.02 \\
                         & BM3D   & LOUPE Unconstrained    & 0.00014 $\pm$ 5e-05   & 0.00094 $\pm$ 8e-05   & 0.0088 $\pm$ 0.0015 & 0.0241 $\pm$ 0.0011 & 0.2972 $\pm$ 0.049  & 0.5298 $\pm$ 0.0845 & 38.81 $\pm$ 1.49 & 30.28 $\pm$ 0.39 & 0.96 $\pm$ 0.01 & 0.81 $\pm$ 0.02 \\ \cline{2-13} 
                         & LORAKS & Cartesian              & 0.22717 $\pm$ 0.25121 & 0.26113 $\pm$ 0.13396 & 0.1785 $\pm$ 0.1043 & 0.1896 $\pm$ 0.0521 & 2.8759 $\pm$ 1.0952 & 3.5001 $\pm$ 0.9479 & 8.45 $\pm$ 3.86  & 6.93 $\pm$ 3.64  & 0.46 $\pm$ 0.07 & 0.28 $\pm$ 0.06 \\
                         & LORAKS & LOUPE Line Constrained & 0.00025 $\pm$ 7e-05   & 0.00097 $\pm$ 0.0001  & 0.0112 $\pm$ 0.0018 & 0.0246 $\pm$ 0.0013 & 0.5801 $\pm$ 0.0295 & 0.7668 $\pm$ 0.0788 & 36.32 $\pm$ 1.41 & 30.14 $\pm$ 0.43 & 0.94 $\pm$ 0.02 & 0.8 $\pm$ 0.02  \\
                         & LORAKS & Uniform Random         & 0.01692 $\pm$ 0.00581 & 0.01861 $\pm$ 0.0002  & 0.0913 $\pm$ 0.0231 & 0.1052 $\pm$ 0.0062 & 1.1886 $\pm$ 0.0497 & 1.2677 $\pm$ 0.0434 & 17.99 $\pm$ 1.57 & 17.3 $\pm$ 0.05  & 0.42 $\pm$ 0.08 & 0.26 $\pm$ 0.02 \\
                         & LORAKS & Variable Density       & 0.00041 $\pm$ 0.00014 & 0.00231 $\pm$ 0.00028 & 0.0152 $\pm$ 0.0025 & 0.0372 $\pm$ 0.0024 & 0.6364 $\pm$ 0.0881 & 1.1334 $\pm$ 0.1639 & 34.1 $\pm$ 1.45  & 26.4 $\pm$ 0.57  & 0.89 $\pm$ 0.04 & 0.64 $\pm$ 0.04 \\
                         & LORAKS & Spectrum               & 0.00036 $\pm$ 0.00013 & 0.00208 $\pm$ 0.00021 & 0.0142 $\pm$ 0.0024 & 0.0355 $\pm$ 0.0017 & 0.6691 $\pm$ 0.0743 & 1.184 $\pm$ 0.1094  & 34.69 $\pm$ 1.45 & 26.85 $\pm$ 0.43 & 0.91 $\pm$ 0.03 & 0.67 $\pm$ 0.03 \\
                         & LORAKS & LOUPE Unconstrained    & 0.00019 $\pm$ 6e-05   & 0.00113 $\pm$ 9e-05   & 0.0105 $\pm$ 0.0017 & 0.0266 $\pm$ 0.0011 & 0.3754 $\pm$ 0.0745 & 0.6854 $\pm$ 0.0853 & 37.42 $\pm$ 1.37 & 29.48 $\pm$ 0.35 & 0.94 $\pm$ 0.02 & 0.78 $\pm$ 0.02 \\ \cline{2-13} 
                         & TV     & Cartesian              & 0.00789 $\pm$ 0.00176 & 0.00582 $\pm$ 0.00129 & 0.0643 $\pm$ 0.0092 & 0.0564 $\pm$ 0.0055 & 0.8938 $\pm$ 0.0028 & 0.9126 $\pm$ 0.0141 & 21.14 $\pm$ 1.01 & 22.46 $\pm$ 0.99 & 0.55 $\pm$ 0.03 & 0.55 $\pm$ 0.03 \\
                         & TV     & LOUPE Line Constrained & 0.00089 $\pm$ 0.00012 & 0.0012 $\pm$ 2e-05    & 0.0217 $\pm$ 0.0019 & 0.027 $\pm$ 0.0002  & 0.9029 $\pm$ 0.1649 & 0.7823 $\pm$ 0.0636 & 30.57 $\pm$ 0.6  & 29.22 $\pm$ 0.07 & 0.8 $\pm$ 0.03  & 0.73 $\pm$ 0.02 \\
                         & TV     & Uniform Random         & 0.01286 $\pm$ 0.00505 & 0.01178 $\pm$ 0.0017  & 0.0755 $\pm$ 0.0229 & 0.0818 $\pm$ 0.01   & 0.9303 $\pm$ 0.0111 & 0.9184 $\pm$ 0.0134 & 19.33 $\pm$ 2.01 & 19.33 $\pm$ 0.61 & 0.62 $\pm$ 0.08 & 0.55 $\pm$ 0.01 \\
                         & TV     & Variable Density       & 0.003 $\pm$ 0.00037   & 0.00333 $\pm$ 0.00013 & 0.0417 $\pm$ 0.0041 & 0.0454 $\pm$ 0.0012 & 1.2104 $\pm$ 0.2833 & 1.0221 $\pm$ 0.0692 & 25.26 $\pm$ 0.54 & 24.78 $\pm$ 0.18 & 0.6 $\pm$ 0.04  & 0.56 $\pm$ 0.06 \\
                         & TV     & Spectrum               & 0.00144 $\pm$ 0.00025 & 0.00188 $\pm$ 6e-05   & 0.0289 $\pm$ 0.0032 & 0.0344 $\pm$ 0.0007 & 0.8695 $\pm$ 0.1871 & 0.8492 $\pm$ 0.0663 & 28.48 $\pm$ 0.72 & 27.27 $\pm$ 0.14 & 0.71 $\pm$ 0.06 & 0.62 $\pm$ 0.02 \\
                         & TV     & LOUPE Unconstrained    & 0.00196 $\pm$ 0.00032 & 0.00218 $\pm$ 0.00011 & 0.0332 $\pm$ 0.0037 & 0.0366 $\pm$ 0.0011 & 0.8991 $\pm$ 0.2634 & 0.7559 $\pm$ 0.0998 & 27.13 $\pm$ 0.75 & 26.63 $\pm$ 0.22 & 0.71 $\pm$ 0.03 & 0.65 $\pm$ 0.02 \\ \cline{2-13} 
                         & U-Net  & Cartesian              & 0.00406 $\pm$ 0.00112 & 0.00345 $\pm$ 0.00038 & 0.0381 $\pm$ 0.0081 & 0.0401 $\pm$ 0.0018 & 0.8265 $\pm$ 0.0395 & 0.8648 $\pm$ 0.0182 & 24.11 $\pm$ 1.37 & 24.65 $\pm$ 0.47 & 0.83 $\pm$ 0.04 & 0.79 $\pm$ 0.01 \\
                         & U-Net  & LOUPE Line Constrained & 0.00015 $\pm$ 5e-05   & 0.00058 $\pm$ 3e-05   & 0.0084 $\pm$ 0.0014 & 0.0186 $\pm$ 0.0004 & 0.3563 $\pm$ 0.0148 & 0.4567 $\pm$ 0.018  & 38.53 $\pm$ 1.52 & 32.36 $\pm$ 0.19 & 0.97 $\pm$ 0.01 & 0.89 $\pm$ 0.0  \\
                         & U-Net  & Uniform Random         & 0.0022 $\pm$ 0.00078  & 0.00204 $\pm$ 0.00017 & 0.0303 $\pm$ 0.0078 & 0.0323 $\pm$ 0.0009 & 0.8342 $\pm$ 0.0974 & 0.7689 $\pm$ 0.0149 & 26.89 $\pm$ 1.7  & 26.93 $\pm$ 0.35 & 0.87 $\pm$ 0.03 & 0.85 $\pm$ 0.01 \\
                         & U-Net  & Variable Density       & 0.00014 $\pm$ 4e-05   & 0.00053 $\pm$ 3e-05   & 0.0086 $\pm$ 0.0015 & 0.018 $\pm$ 0.0005  & 0.4088 $\pm$ 0.0231 & 0.5186 $\pm$ 0.0158 & 38.79 $\pm$ 1.5  & 32.74 $\pm$ 0.21 & 0.97 $\pm$ 0.01 & 0.9 $\pm$ 0.0   \\
                         & U-Net  & Spectrum               & 0.00013 $\pm$ 4e-05   & 0.00053 $\pm$ 3e-05   & 0.0084 $\pm$ 0.0014 & 0.0179 $\pm$ 0.0005 & 0.4266 $\pm$ 0.02   & 0.5727 $\pm$ 0.019  & 39.05 $\pm$ 1.52 & 32.77 $\pm$ 0.22 & 0.97 $\pm$ 0.01 & 0.9 $\pm$ 0.0   \\
                         & U-Net  & LOUPE Unconstrained    & 9e-05 $\pm$ 3e-05     & 0.00048 $\pm$ 1e-05   & 0.0071 $\pm$ 0.0011 & 0.0173 $\pm$ 0.0002 & 0.1681 $\pm$ 0.01   & 0.246 $\pm$ 0.0162  & 40.71 $\pm$ 1.44 & 33.16 $\pm$ 0.09 & 0.98 $\pm$ 0.01 & 0.91 $\pm$ 0.0  \\ \hline
\multirow{24}{*}{8-fold} & BM3D   & Cartesian              & 0.00871 $\pm$ 0.00189 & 0.00671 $\pm$ 0.00146 & 0.0665 $\pm$ 0.0094 & 0.0603 $\pm$ 0.0054 & 0.9855 $\pm$ 0.0069 & 1.0604 $\pm$ 0.012  & 20.71 $\pm$ 0.99 & 21.84 $\pm$ 0.95 & 0.53 $\pm$ 0.02 & 0.6 $\pm$ 0.02  \\
                         & BM3D   & LOUPE Line Constrained & 0.00059 $\pm$ 0.00024 & 0.00124 $\pm$ 0.0001  & 0.0152 $\pm$ 0.0029 & 0.0273 $\pm$ 0.0012 & 0.7843 $\pm$ 0.0395 & 0.8844 $\pm$ 0.0559 & 32.66 $\pm$ 1.7  & 29.08 $\pm$ 0.36 & 0.9 $\pm$ 0.02  & 0.75 $\pm$ 0.02 \\
                         & BM3D   & Uniform Random         & 0.01248 $\pm$ 0.00567 & 0.00854 $\pm$ 0.00211 & 0.0759 $\pm$ 0.024  & 0.0683 $\pm$ 0.0112 & 1.0331 $\pm$ 0.0744 & 1.1762 $\pm$ 0.0649 & 19.56 $\pm$ 2.22 & 20.81 $\pm$ 0.98 & 0.59 $\pm$ 0.09 & 0.5 $\pm$ 0.04  \\
                         & BM3D   & Variable Density       & 0.00097 $\pm$ 0.00038 & 0.00153 $\pm$ 0.00012 & 0.0217 $\pm$ 0.0045 & 0.0306 $\pm$ 0.0011 & 0.706 $\pm$ 0.0639  & 0.9147 $\pm$ 0.0872 & 30.45 $\pm$ 1.6  & 28.16 $\pm$ 0.34 & 0.87 $\pm$ 0.03 & 0.73 $\pm$ 0.03 \\
                         & BM3D   & Spectrum               & 0.00032 $\pm$ 0.00012 & 0.00148 $\pm$ 0.00012 & 0.0128 $\pm$ 0.0023 & 0.0301 $\pm$ 0.0013 & 0.6758 $\pm$ 0.0463 & 0.9278 $\pm$ 0.071  & 35.26 $\pm$ 1.65 & 28.32 $\pm$ 0.36 & 0.92 $\pm$ 0.02 & 0.7 $\pm$ 0.03  \\
                         & BM3D   & LOUPE Unconstrained    & 0.0002 $\pm$ 7e-05    & 0.00104 $\pm$ 9e-05   & 0.0103 $\pm$ 0.0017 & 0.0256 $\pm$ 0.0012 & 0.4684 $\pm$ 0.0444 & 0.7022 $\pm$ 0.0835 & 37.16 $\pm$ 1.52 & 29.85 $\pm$ 0.39 & 0.93 $\pm$ 0.02 & 0.76 $\pm$ 0.03 \\ \cline{2-13} 
                         & LORAKS & Cartesian              & 0.4987 $\pm$ 0.32268  & 0.34232 $\pm$ 0.07458 & 0.2875 $\pm$ 0.0871 & 0.2739 $\pm$ 0.0318 & 4.2092 $\pm$ 1.071  & 4.1481 $\pm$ 0.9071 & 4.05 $\pm$ 3.22  & 4.77 $\pm$ 0.99  & 0.44 $\pm$ 0.06 & 0.31 $\pm$ 0.03 \\
                         & LORAKS & LOUPE Line Constrained & 0.00046 $\pm$ 0.00013 & 0.00138 $\pm$ 0.00011 & 0.0147 $\pm$ 0.002  & 0.029 $\pm$ 0.0013  & 0.8339 $\pm$ 0.0421 & 0.9979 $\pm$ 0.0701 & 33.56 $\pm$ 1.27 & 28.63 $\pm$ 0.36 & 0.9 $\pm$ 0.02  & 0.73 $\pm$ 0.02 \\
                         & LORAKS & Uniform Random         & 1.15118 $\pm$ 0.80465 & 0.86483 $\pm$ 0.21083 & 0.3866 $\pm$ 0.1535 & 0.3221 $\pm$ 0.0275 & 9.7904 $\pm$ 3.022  & 6.5299 $\pm$ 0.6076 & 0.26 $\pm$ 2.62  & 0.77 $\pm$ 1.13  & 0.35 $\pm$ 0.07 & 0.17 $\pm$ 0.03 \\
                         & LORAKS & Variable Density       & 0.00041 $\pm$ 0.00013 & 0.00188 $\pm$ 0.00014 & 0.0152 $\pm$ 0.0025 & 0.0343 $\pm$ 0.0014 & 0.7671 $\pm$ 0.0731 & 1.2851 $\pm$ 0.1822 & 34.12 $\pm$ 1.5  & 27.27 $\pm$ 0.34 & 0.89 $\pm$ 0.03 & 0.65 $\pm$ 0.04 \\
                         & LORAKS & Spectrum               & 0.00038 $\pm$ 0.00011 & 0.00175 $\pm$ 0.00015 & 0.0145 $\pm$ 0.0023 & 0.033 $\pm$ 0.0015  & 0.7555 $\pm$ 0.0334 & 1.1352 $\pm$ 0.0907 & 34.45 $\pm$ 1.36 & 27.59 $\pm$ 0.37 & 0.9 $\pm$ 0.03  & 0.66 $\pm$ 0.03 \\
                         & LORAKS & LOUPE Unconstrained    & 0.00022 $\pm$ 7e-05   & 0.00111 $\pm$ 0.00011 & 0.0111 $\pm$ 0.0018 & 0.0266 $\pm$ 0.0014 & 0.5038 $\pm$ 0.054  & 0.8037 $\pm$ 0.0823 & 36.79 $\pm$ 1.41 & 29.56 $\pm$ 0.45 & 0.92 $\pm$ 0.02 & 0.75 $\pm$ 0.02 \\ \cline{2-13} 
                         & TV     & Cartesian              & 0.0086 $\pm$ 0.00189  & 0.00631 $\pm$ 0.00143 & 0.0664 $\pm$ 0.0094 & 0.0585 $\pm$ 0.0055 & 0.9508 $\pm$ 0.0016 & 0.9606 $\pm$ 0.008  & 20.76 $\pm$ 1.0  & 22.11 $\pm$ 0.99 & 0.46 $\pm$ 0.04 & 0.45 $\pm$ 0.06 \\
                         & TV     & LOUPE Line Constrained & 0.00114 $\pm$ 0.00018 & 0.00153 $\pm$ 5e-05   & 0.0245 $\pm$ 0.0021 & 0.0302 $\pm$ 0.0004 & 1.1597 $\pm$ 0.244  & 1.0106 $\pm$ 0.058  & 29.48 $\pm$ 0.67 & 28.15 $\pm$ 0.15 & 0.79 $\pm$ 0.03 & 0.71 $\pm$ 0.02 \\
                         & TV     & Uniform Random         & 0.01392 $\pm$ 0.00634 & 0.01097 $\pm$ 0.00251 & 0.0792 $\pm$ 0.0274 & 0.0785 $\pm$ 0.0128 & 0.975 $\pm$ 0.0079  & 0.9729 $\pm$ 0.0058 & 19.15 $\pm$ 2.41 & 19.7 $\pm$ 0.91  & 0.6 $\pm$ 0.08  & 0.55 $\pm$ 0.03 \\
                         & TV     & Variable Density       & 0.00286 $\pm$ 0.00043 & 0.0032 $\pm$ 0.00012  & 0.0414 $\pm$ 0.0043 & 0.0445 $\pm$ 0.0011 & 1.2487 $\pm$ 0.3041 & 1.076 $\pm$ 0.0697  & 25.48 $\pm$ 0.69 & 24.95 $\pm$ 0.17 & 0.58 $\pm$ 0.05 & 0.56 $\pm$ 0.07 \\
                         & TV     & Spectrum               & 0.00199 $\pm$ 0.00032 & 0.0024 $\pm$ 9e-05    & 0.034 $\pm$ 0.0036  & 0.0388 $\pm$ 0.0008 & 1.0703 $\pm$ 0.1989 & 0.9868 $\pm$ 0.0546 & 27.07 $\pm$ 0.73 & 26.21 $\pm$ 0.17 & 0.66 $\pm$ 0.04 & 0.59 $\pm$ 0.07 \\
                         & TV     & LOUPE Unconstrained    & 0.00229 $\pm$ 0.00031 & 0.0027 $\pm$ 0.00012  & 0.0364 $\pm$ 0.0036 & 0.0407 $\pm$ 0.0011 & 1.2016 $\pm$ 0.3756 & 1.0351 $\pm$ 0.1059 & 26.44 $\pm$ 0.61 & 25.69 $\pm$ 0.19 & 0.66 $\pm$ 0.04 & 0.57 $\pm$ 0.01 \\ \cline{2-13} 
                         & U-Net  & Cartesian              & 0.007 $\pm$ 0.00191   & 0.00509 $\pm$ 0.00075 & 0.0517 $\pm$ 0.0101 & 0.0489 $\pm$ 0.0023 & 1.0244 $\pm$ 0.0283 & 0.988 $\pm$ 0.0041  & 21.72 $\pm$ 1.26 & 22.97 $\pm$ 0.6  & 0.73 $\pm$ 0.02 & 0.68 $\pm$ 0.02 \\
                         & U-Net  & LOUPE Line Constrained & 0.00027 $\pm$ 9e-05   & 0.00079 $\pm$ 6e-05   & 0.0108 $\pm$ 0.0019 & 0.0214 $\pm$ 0.0007 & 0.6327 $\pm$ 0.0499 & 0.7301 $\pm$ 0.0225 & 36.0 $\pm$ 1.53  & 31.01 $\pm$ 0.3  & 0.95 $\pm$ 0.01 & 0.85 $\pm$ 0.01 \\
                         & U-Net  & Uniform Random         & 0.00316 $\pm$ 0.00089 & 0.00279 $\pm$ 0.00026 & 0.0363 $\pm$ 0.0081 & 0.0376 $\pm$ 0.0022 & 0.8189 $\pm$ 0.0472 & 0.8267 $\pm$ 0.0133 & 25.17 $\pm$ 1.24 & 25.56 $\pm$ 0.39 & 0.84 $\pm$ 0.03 & 0.8 $\pm$ 0.02  \\
                         & U-Net  & Variable Density       & 0.00027 $\pm$ 8e-05   & 0.00074 $\pm$ 4e-05   & 0.0118 $\pm$ 0.0019 & 0.0211 $\pm$ 0.0006 & 0.4746 $\pm$ 0.0268 & 0.5759 $\pm$ 0.0193 & 35.82 $\pm$ 1.31 & 31.32 $\pm$ 0.24 & 0.96 $\pm$ 0.01 & 0.87 $\pm$ 0.0  \\
                         & U-Net  & Spectrum               & 0.00023 $\pm$ 8e-05   & 0.00075 $\pm$ 5e-05   & 0.0105 $\pm$ 0.0019 & 0.021 $\pm$ 0.0007  & 0.5828 $\pm$ 0.0394 & 0.7233 $\pm$ 0.0199 & 36.69 $\pm$ 1.59 & 31.28 $\pm$ 0.29 & 0.96 $\pm$ 0.01 & 0.86 $\pm$ 0.01 \\
                         & U-Net  & LOUPE Unconstrained    & 0.00014 $\pm$ 5e-05   & 0.00066 $\pm$ 2e-05   & 0.0085 $\pm$ 0.0014 & 0.0199 $\pm$ 0.0003 & 0.3044 $\pm$ 0.0175 & 0.4303 $\pm$ 0.0229 & 38.85 $\pm$ 1.59 & 31.83 $\pm$ 0.15 & 0.97 $\pm$ 0.01 & 0.87 $\pm$ 0.0 
\end{tabular}
\end{sidewaystable*}

\end{document}